\documentclass[12pt]{article}
\usepackage{graphicx}
\usepackage{ifthen}
\usepackage{xspace}
\usepackage{upgreek}

\newboolean{uprightparticles}
\setboolean{uprightparticles}{false}

\newcommand\pubnumber{DPF2013-148}
\newcommand\pubdate{\today}

\def\warwick{Department of Physics, University of Warwick,\\
Coventry, CV4 7AL, United Kingdom}
\def\onbehalfof{\footnote{On behalf of the LHCb collaboration.}}

\def\Title#1{\begin{center} {\Large #1 } \end{center}}
\def\Author#1{\begin{center}{ \sc #1} \end{center}}
\def\Address#1{\begin{center}{ \it #1} \end{center}}

\newcommand\pubblock{\rightline{\begin{tabular}{l} \pubnumber\\
         \pubdate  \end{tabular}}}
\newenvironment{Abstract}{\begin{quotation}  }{\end{quotation}}
\newenvironment{Presented}{\begin{quotation} \begin{center} 
             PRESENTED AT\end{center}\bigskip 
      \begin{center}\begin{large}}{\end{large}\end{center} \end{quotation}}
\def\Acknowledgments{\bigskip  \bigskip \begin{center} \begin{large}
             \bf ACKNOWLEDGMENTS \end{large}\end{center}}
\def\beq{\begin{equation}}
\def\eeq#1{\label{#1}\end{equation}}
\def\eeqn{\end{equation}}

\def\beqa{\begin{eqnarray}}
\def\eeqa#1{\label{#1}\end{eqnarray}}
\def\eeqan{\end{eqnarray}}

\let\bar=\overbar

\def\D{{\cal D}}

\def\Dslash{\not{\hbox{\kern-4pt $D$}}}
\def\dslash{\not{\hbox{\kern-2pt $\del$}}}

\def\msb{{\bar{\ssstyle M \kern -1pt S}}}

\def\babar  {\mbox{BaBar}\xspace}
\def\belle  {\mbox{Belle}\xspace}

\ifthenelse{\boolean{uprightparticles}}%
{

 \def\Pmu         {\ensuremath{\upmu}\xspace}

 \def\Ppi         {\ensuremath{\uppi}\xspace}                 
                  
 \def\Prho        {\ensuremath{\uprho}\xspace}

 \def\Pphi        {\ensuremath{\upphi}\xspace}

 \def\Ppsi        {\ensuremath{\uppsi}\xspace}

 \def\PDelta      {\ensuremath{\Delta}\xspace}                 
 \def\PXi      {\ensuremath{\Xi}\xspace}                 
 \def\PLambda      {\ensuremath{\Lambda}\xspace}                 
 \def\PSigma      {\ensuremath{\Sigma}\xspace}                 
 \def\POmega      {\ensuremath{\Omega}\xspace}                 
 \def\PUpsilon      {\ensuremath{\Upsilon}\xspace}                 
 
 \def\PB      {\ensuremath{\mathrm{B}}\xspace}                 
                  
 \def\PD      {\ensuremath{\mathrm{D}}\xspace}

 \def\PJ      {\ensuremath{\mathrm{J}}\xspace}                 
 \def\PK      {\ensuremath{\mathrm{K}}\xspace}

 \def\Pb      {\ensuremath{\mathrm{b}}\xspace}                 
                  
 \def\Pd      {\ensuremath{\mathrm{d}}\xspace}

 \def\Ph      {\ensuremath{\mathrm{h}}\xspace}                 
 \def\Pi      {\ensuremath{\mathrm{i}}\xspace}

 \def\Pp      {\ensuremath{\mathrm{p}}\xspace}

 \def\Ps      {\ensuremath{\mathrm{s}}\xspace}                 
                  
 \def\Pu      {\ensuremath{\mathrm{u}}\xspace}

}
{

 \def\Pmu         {\ensuremath{\mu}\xspace}

 \def\Ppi         {\ensuremath{\pi}\xspace}                 
                  
 \def\Prho        {\ensuremath{\rho}\xspace}

 \def\Pphi        {\ensuremath{\phi}\xspace}

 \def\Ppsi        {\ensuremath{\psi}\xspace}                 
                  
 \mathchardef\PDelta="7101
 \mathchardef\PXi="7104
 \mathchardef\PLambda="7103
 \mathchardef\PSigma="7106
 \mathchardef\POmega="710A
 \mathchardef\PUpsilon="7107
                  
 \def\PB      {\ensuremath{B}\xspace}                 
                  
 \def\PD      {\ensuremath{D}\xspace}

 \def\PJ      {\ensuremath{J}\xspace}                 
 \def\PK      {\ensuremath{K}\xspace}

 \def\Pb      {\ensuremath{b}\xspace}                 
                  
 \def\Pd      {\ensuremath{d}\xspace}

 \def\Ph      {\ensuremath{h}\xspace}                 
 \def\Pi      {\ensuremath{i}\xspace}

 \def\Pp      {\ensuremath{p}\xspace}

 \def\Ps      {\ensuremath{s}\xspace}                 
                  
 \def\Pu      {\ensuremath{u}\xspace}

}

\def\mumu       {\ensuremath{\Pmu^+\Pmu^-}\xspace}

\def\uquark    {\ensuremath{\Pu}\xspace}

\def\dquark    {\ensuremath{\Pd}\xspace}

\def\squark    {\ensuremath{\Ps}\xspace}

\def\bquark    {\ensuremath{\Pb}\xspace}

\def\pion  {\ensuremath{\Ppi}\xspace}
\def\piz   {\ensuremath{\pion^0}\xspace}

\def\pip   {\ensuremath{\pion^+}\xspace}
\def\pim   {\ensuremath{\pion^-}\xspace}
\def\pipi  {\ensuremath{\pion^+\pion^-}\xspace}

\def\pimp  {\ensuremath{\pion^\mp}\xspace}

\def\kaon  {\ensuremath{\PK}\xspace}
  \def\Kbar  {\kern 0.2em\overline{\kern -0.2em \PK}{}\xspace}

\def\Kz    {\ensuremath{\kaon^0}\xspace}
\def\Kzb   {\ensuremath{\Kbar^0}\xspace}
\def\KzKzb {\ensuremath{\Kz \kern -0.16em \Kzb}\xspace}
\def\Kp    {\ensuremath{\kaon^+}\xspace}
\def\Km    {\ensuremath{\kaon^-}\xspace}
\def\Kpm   {\ensuremath{\kaon^\pm}\xspace}

\def\KpKm  {\ensuremath{\Kp \kern -0.16em \Km}\xspace}
\def\KS    {\ensuremath{\kaon^0_{\rm\scriptscriptstyle S}}\xspace}

  \def\Dbar    {\kern 0.2em\overline{\kern -0.2em \PD}{}\xspace}
\def\D       {\ensuremath{\PD}\xspace}

\def\Dz      {\ensuremath{\D^0}\xspace}
\def\Dzb     {\ensuremath{\Dbar^0}\xspace}
\def\DzDzb   {\ensuremath{\Dz {\kern -0.16em \Dzb}}\xspace}
\def\Dp      {\ensuremath{\D^+}\xspace}
\def\Dm      {\ensuremath{\D^-}\xspace}

\def\DpDm    {\ensuremath{\Dp {\kern -0.16em \Dm}}\xspace}

\def\B       {\ensuremath{\PB}\xspace}
\def\Bbar    {\ensuremath{\kern 0.18em\overline{\kern -0.18em \PB}{}}\xspace}
\def\Bb      {\ensuremath{\Bbar}\xspace}
\def\BBbar   {\ensuremath{\B\Bbar}\xspace} 
\def\Bz      {\ensuremath{\B^0}\xspace}

\def\Bu      {\ensuremath{\B^+}\xspace}
\def\Bub     {\ensuremath{\B^-}\xspace}
\def\Bp      {\ensuremath{\Bu}\xspace}
\def\Bm      {\ensuremath{\Bub}\xspace}
\def\Bpm     {\ensuremath{\B^\pm}\xspace}

\def\Bd      {\ensuremath{\B^0}\xspace}
\def\Bs      {\ensuremath{\B^0_\squark}\xspace}

\def\Bdb     {\ensuremath{\Bbar^0}\xspace}

\def\Bds     {\ensuremath{\B^0_{(\squark)}}\xspace}

\def\fbar {\ensuremath{\kern 0.18em\overline{\kern -0.18em f}{}}\xspace}

\def\jpsi     {\ensuremath{{\PJ\mskip -3mu/\mskip -2mu\Ppsi\mskip 2mu}}\xspace}

  \def\Y#1S{\ensuremath{\PUpsilon{(#1S)}}\xspace}% no space before {...}!

\def\FourS {\Y4S}

\def\proton      {\ensuremath{\Pp}\xspace}
\def\antiproton  {\ensuremath{\overline \proton}\xspace}

\def\pp          {\ensuremath{\proton\proton}\xspace}
\def\ppbar       {\ensuremath{\proton\antiproton}\xspace}

\def\Lz {\ensuremath{\PLambda}\xspace}
\def\Lbar {\ensuremath{\kern 0.1em\overline{\kern -0.1em\PLambda}}\xspace}

\def\Lb      {\ensuremath{\Lz^0_\bquark}\xspace}

\def\BF         {{\ensuremath{\cal B}\xspace}}

\newcommand{\Br}[1]{\ensuremath{\BF\left(#1\right)}\xspace}
\newcommand{\decay}[2]{\ensuremath{#1\!\to #2}\xspace}         % {\Pa}{\Pb \Pc}

\def\to                 {\ensuremath{\rightarrow}\xspace}

\newcommand{\tauBs}{\ensuremath{\tau_{\Bs}}\xspace}
\newcommand{\tauBd}{\ensuremath{\tau_{\Bd}}\xspace}

\def\CP                {\ensuremath{C\!P}\xspace}

\newcommand{\ACP}{\ensuremath{{\cal A}^{\CP}}\xspace}

\def\BToKpi       {\decay{\B}{\kaon\pion}}

\def\BdToKpi      {\decay{\Bd}{\Kp\pim}}

\def\BsTopiK      {\decay{\Bs}{\pip\Km}}

\def\rhoz         {\ensuremath{\Prho^0}\xspace}
\def\rhop         {\ensuremath{\Prho^+}\xspace}
\def\rhom         {\ensuremath{\Prho^-}\xspace}

\def\BuTohhh      {\decay{\Bp}{\Ph^+ \Ph^+ \Ph^-}}
\def\BuTorhoK     {\decay{\Bp}{\rhoz(770) \Kp}}
\def\BuTophiK     {\decay{\Bp}{\Pphi(1020) \Kp}}
\def\BuToKhh      {\decay{\Bp}{\Kp \Ph^+ \Ph^-}}
\def\BuToJPsiK    {\decay{\Bp}{\jpsi \Kp}}
\def\BuTopipipi   {\decay{\Bp}{\pip \pip \pim}}
\def\BuToKpipi    {\decay{\Bp}{\Kp \pip \pim}}
\def\BuToKKpi     {\decay{\Bp}{\pip \Kp \Km}}
\def\BuToKKK      {\decay{\Bp}{\Kp \Kp \Km}}

\def\BuTopph      {\decay{\Bp}{\proton \antiproton \Ph^+}}
\def\BuTopppi     {\decay{\Bp}{\proton \antiproton \pip}}
\def\BuToppK      {\decay{\Bp}{\proton \antiproton \Kp}}

\def\Kproton      {\ensuremath{\Kp\antiproton}\xspace}
\def\BuToLbp      {\decay{\Bp}{\Lbar(1520)\proton}}
\def\JPsiTopp     {\decay{\jpsi}{\ppbar}}

\def\BdTohhpiz    {\decay{\Bd}{\Ph^+ \Ph^- \piz}}
\def\BdTopipipiz  {\decay{\Bd}{\pip \pim \piz}}
\def\BdToKKpiz    {\decay{\Bd}{\Kp \Km \piz}}

\def\BdTorhoppim  {\decay{\Bd}{\rhop\pim}}
\def\BdbTorhoppim {\decay{\Bdb}{\rhop\pim}}
\def\BdTorhompip  {\decay{\Bd}{\rhom\pip}}
\def\BdbTorhompip {\decay{\Bdb}{\rhom\pip}}

\def\KShh         {\ensuremath{\KS \Ph^+ \Ph^-}\xspace}
\def\KSpipi       {\ensuremath{\KS \pip \pim}\xspace}
\def\KSKpi        {\ensuremath{\KS \Kpm \pimp}\xspace}
\def\KSKK         {\ensuremath{\KS \Kp \Km}\xspace}

\def\BdsToKShh    {\decay{\Bds}{\KS \Ph^+ \Ph^-}}

\def\BdToKSpipi   {\decay{\Bd}{\KS \pip \pim}}
\def\BsToKSpipi   {\decay{\Bs}{\KS \pip \pim}}

\def\BdToKSKpi    {\decay{\Bd}{\KS \Kpm \pimp}}
\def\BsToKSKpi    {\decay{\Bs}{\KS \Kpm \pimp}}

\def\BdToKSKK     {\decay{\Bd}{\KS \Kp \Km}}
\def\BsToKSKK     {\decay{\Bs}{\KS \Kp \Km}}

\def\AFB      {\ensuremath{A_{\mathrm{FB}}}\xspace}

\def\AT#1     {\ensuremath{A_{\mathrm{T}}^{#1}}\xspace}           % 2

\def\C#1      {\ensuremath{\mathcal{C}_{#1}}\xspace}                       % 9
\def\Cp#1     {\ensuremath{\mathcal{C}_{#1}^{'}}\xspace}                    % 7
\def\Ceff#1   {\ensuremath{\mathcal{C}_{#1}^{\mathrm{(eff)}}}\xspace}        % 9  
\def\Cpeff#1  {\ensuremath{\mathcal{C}_{#1}^{'\mathrm{(eff)}}}\xspace}       % 7
\def\Ope#1    {\ensuremath{\mathcal{O}_{#1}}\xspace}                       % 2
\def\Opep#1   {\ensuremath{\mathcal{O}_{#1}^{'}}\xspace}                    % 7

\newcommand{\tev}{\ifthenelse{\boolean{inbibliography}}{\ensuremath{~T\kern -0.05em eV}\xspace}{\ensuremath{\mathrm{\,Te\kern -0.1em V}}\xspace}}
\newcommand{\gev}{\ensuremath{\mathrm{\,Ge\kern -0.1em V}}\xspace}
\newcommand{\mev}{\ensuremath{\mathrm{\,Me\kern -0.1em V}}\xspace}
\newcommand{\kev}{\ensuremath{\mathrm{\,ke\kern -0.1em V}}\xspace}
\newcommand{\ev}{\ensuremath{\mathrm{\,e\kern -0.1em V}}\xspace}
\newcommand{\gevc}{\ensuremath{{\mathrm{\,Ge\kern -0.1em V\!/}c}}\xspace}
\newcommand{\mevc}{\ensuremath{{\mathrm{\,Me\kern -0.1em V\!/}c}}\xspace}
\newcommand{\gevcc}{\ensuremath{{\mathrm{\,Ge\kern -0.1em V\!/}c^2}}\xspace}
\newcommand{\gevgevcccc}{\ensuremath{{\mathrm{\,Ge\kern -0.1em V^2\!/}c^4}}\xspace}
\newcommand{\mevcc}{\ensuremath{{\mathrm{\,Me\kern -0.1em V\!/}c^2}}\xspace}

\def\invfb   {\ensuremath{\mbox{\,fb}^{-1}}\xspace}

\newcommand{\stat}{\ensuremath{\mathrm{\,(stat)}}\xspace}
\newcommand{\syst}{\ensuremath{\mathrm{\,(syst)}}\xspace}

\newcommand{\chisq}{\ensuremath{\chi^2}\xspace}

\def\gsim{{~\raise.15em\hbox{$>$}\kern-.85em
          \lower.35em\hbox{$\sim$}~}\xspace}
\def\lsim{{~\raise.15em\hbox{$<$}\kern-.85em
          \lower.35em\hbox{$\sim$}~}\xspace}

\def\sPlot{\mbox{\em sPlot}\xspace}

\def\degrees{\ensuremath{^{\circ}}\xspace}

\def\tell1  {TELL1\xspace}
\def\ukl1   {UKL1\xspace}

\begin{document}
\begin{titlepage}
\pubblock

\vfill
\Title{Charmless three-body decays of \bquark-hadrons}
\vfill
\Author{Thomas Latham\onbehalfof}
\Address{\warwick}
\vfill
\begin{Abstract}
A review of recent results from LHCb and the \B-factories on the charmless decays of \bquark-hadrons into three-body final states is presented.
\end{Abstract}
\vfill
\begin{Presented}
DPF 2013\\
The Meeting of the American Physical Society\\
Division of Particles and Fields\\
Santa Cruz, California, August 13--17, 2013\\
\end{Presented}
\vfill
\end{titlepage}
\def\thefootnote{\fnsymbol{footnote}}
\setcounter{footnote}{0}

\section{Introduction}

Charmless decays of \bquark-hadrons can proceed through both
\decay{\bquark}{\uquark} tree and \decay{\bquark}{\squark,\dquark} loop
(penguin) diagrams, which can interfere.
Since they have a relative weak phase of $\gamma$ and the diagrams appear at
similar orders, this can give rise to large direct \CP violation.
In the decays of neutral \B mesons, time-dependent analyses allow measurements
of mixing-induced \CP asymmetries.  Comparing the values of these asymmetries
with those measured in tree-dominated decays such as \decay{\Bz}{\jpsi\KS} or
\decay{\Bs}{\jpsi\Pphi} can be a sensitive test of the Standard Model (SM), with
significant deviations being a sign that new physics particles could be appearing
in the loops.

Recent results from LHCb for the direct \CP asymmetries, defined as
\begin{equation}
\ACP(\decay{\B}{f}) = \frac{\Gamma(\decay{\Bb}{\fbar})-\Gamma(\decay{\B}{f})}{\Gamma(\decay{\Bb}{\fbar})+\Gamma(\decay{\B}{f})} \,,
\end{equation}
of the decays \decay{\Bz}{\Kp\pim} and \decay{\Bs}{\Km\pip}~\cite{Aaij:2013iua}
exhibit large central values\footnote{The inclusion of charge conjugate processes is implied throughout, except in \ACP definitions.}
\begin{eqnarray}
\nonumber
\ACP(\BsTopiK) &=& \phantom{-}0.27\phantom{0} \pm 0.04\phantom{0} \stat \pm 0.01\phantom{0} \syst \,, \\
\nonumber
\ACP(\BdToKpi) &=& -0.080 \pm 0.007 \stat \pm 0.003 \syst \,.
\end{eqnarray}
The first of these constitutes the first observation of \CP violation in the \Bs
system with a significance of $6.5\,\sigma$, while the latter is the world's
most precise single measurement of that quantity.
Combining these results with related quantities in the expression
\begin{eqnarray}
\nonumber
\Delta \equiv \frac{\ACP(\BdToKpi)}{\ACP(\BsTopiK)}
+ \frac{\BF(\BsTopiK)}{\BF(\BdToKpi)} \frac{\tauBd}{\tauBs}
= -0.02 \pm 0.05 \pm 0.04 \,,
\end{eqnarray}
it is found that everything is consistent with the SM expectation
($\Delta=0$)~\cite{Gronau:2000zy}.

It is necessary to form such a combination of quantities in order to test for
compatibility, or otherwise, with the SM because the source of the strong
phase difference is not well understood in two-body decays.  Three-body decays,
on the other hand, allow direct measurements of the relative strong phases
through an amplitude analysis of the Dalitz plot.  Determining both the magnitudes
and the phases of the intermediate states provides greater information for
constraining theoretical models.  In addition, modelling the interferences can
help to resolve trigonometric ambiguities in the measurement of weak phases,
see for example Ref.~\cite{Latham:2008zs}.

\section{\boldmath Direct \CP violation in \BuTohhh decays}

Searches for direct \CP violation in \BuTohhh decays, where $\Ph=\pion,\kaon$ are
motivated both by the large asymmetries seen in \BToKpi decays and \B-factory
results that have shown evidence for direct \CP asymmetries in
\BuTorhoK~\cite{Garmash:2005rv,Aubert:2008bj} and \BuTophiK~\cite{Lees:2012kxa}.
The recent LHCb analysis of \BuToKhh decays makes measurements of the global
\CP asymmetry as well as the local asymmetries in regions of the Dalitz-plot.

The analysis, full details of which can be found in Ref.~\cite{Aaij:2013sfa},
uses the 1.0\invfb of \pp collision data collected during 2011 by the LHCb
detector~\cite{Alves:2008zz}.  The raw asymmetry of measured yields
\begin{equation}
\ACP_{\rm RAW} = \frac{N_{\Bm}-N_{\Bp}}{N_{\Bm}+N_{\Bp}}
\end{equation}
is determined from a simultaneous fit to the sample of \Bp and \Bm candidates.
The raw asymmetry must be corrected for both production and detection asymmetries
\begin{equation}
\ACP = \ACP_{\rm RAW} - {\cal A}_P(\Bpm) - {\cal A}_D(\Kpm) \,,
\end{equation}
which are determined from the control channel \BuToJPsiK, where \jpsi decays to
\mumu, according to the relation
\begin{equation}
{\cal A}_D(\Kpm) + {\cal A}_P(\Bpm) = \ACP_{\rm RAW}(\jpsi\Kp) - \ACP(\jpsi\Kp) \,.
\end{equation}
This channel is well suited for this role due to its similar topology to the
signal channel and since its \CP asymmetry is consistent with zero and precisely
determined, $\ACP(\jpsi\Kp) = (0.1 \pm 0.7)\%$~\cite{Beringer:2012zz}.

\begin{figure}[htb]
\centering
\includegraphics[width=0.69\textwidth]{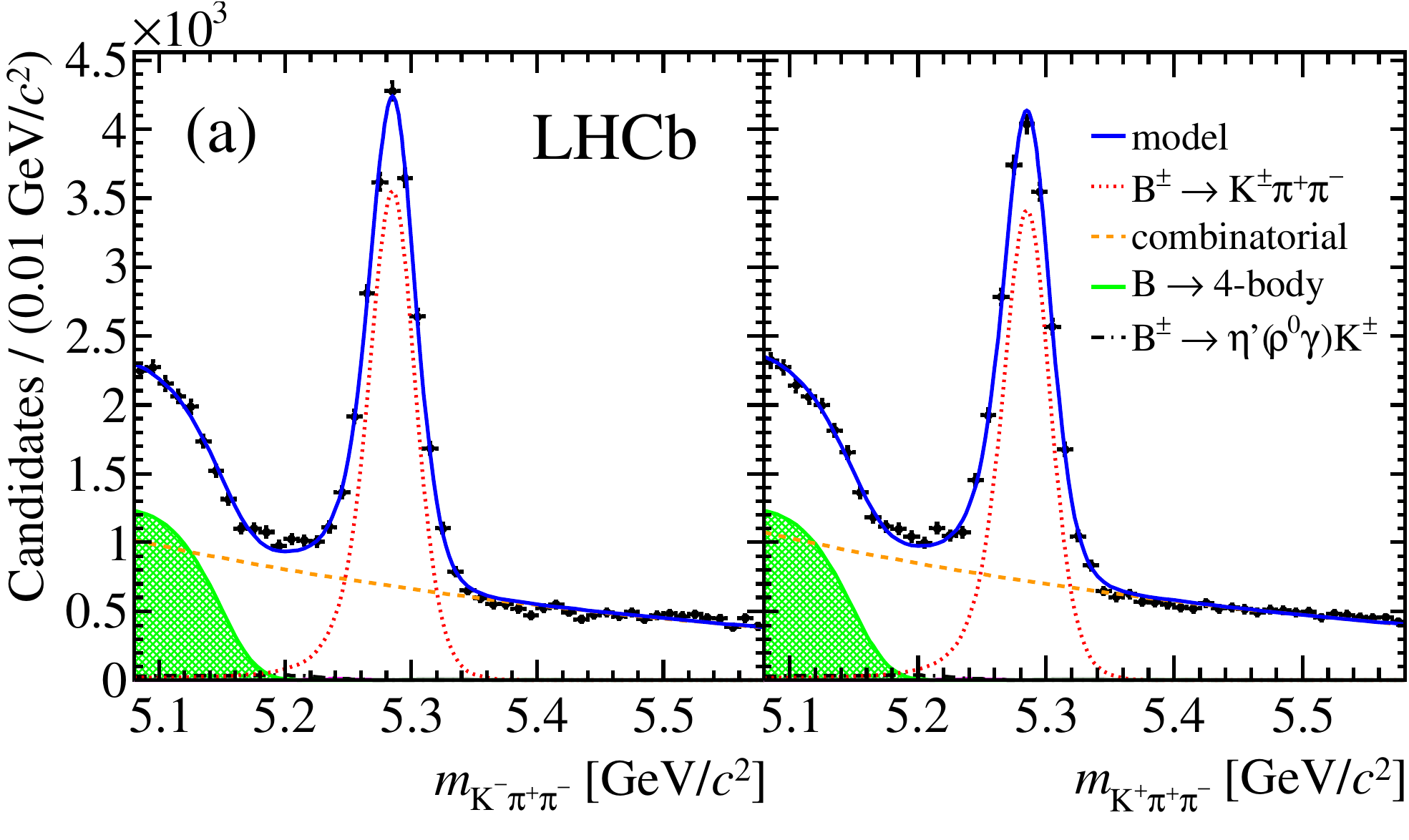}
\includegraphics[width=0.69\textwidth]{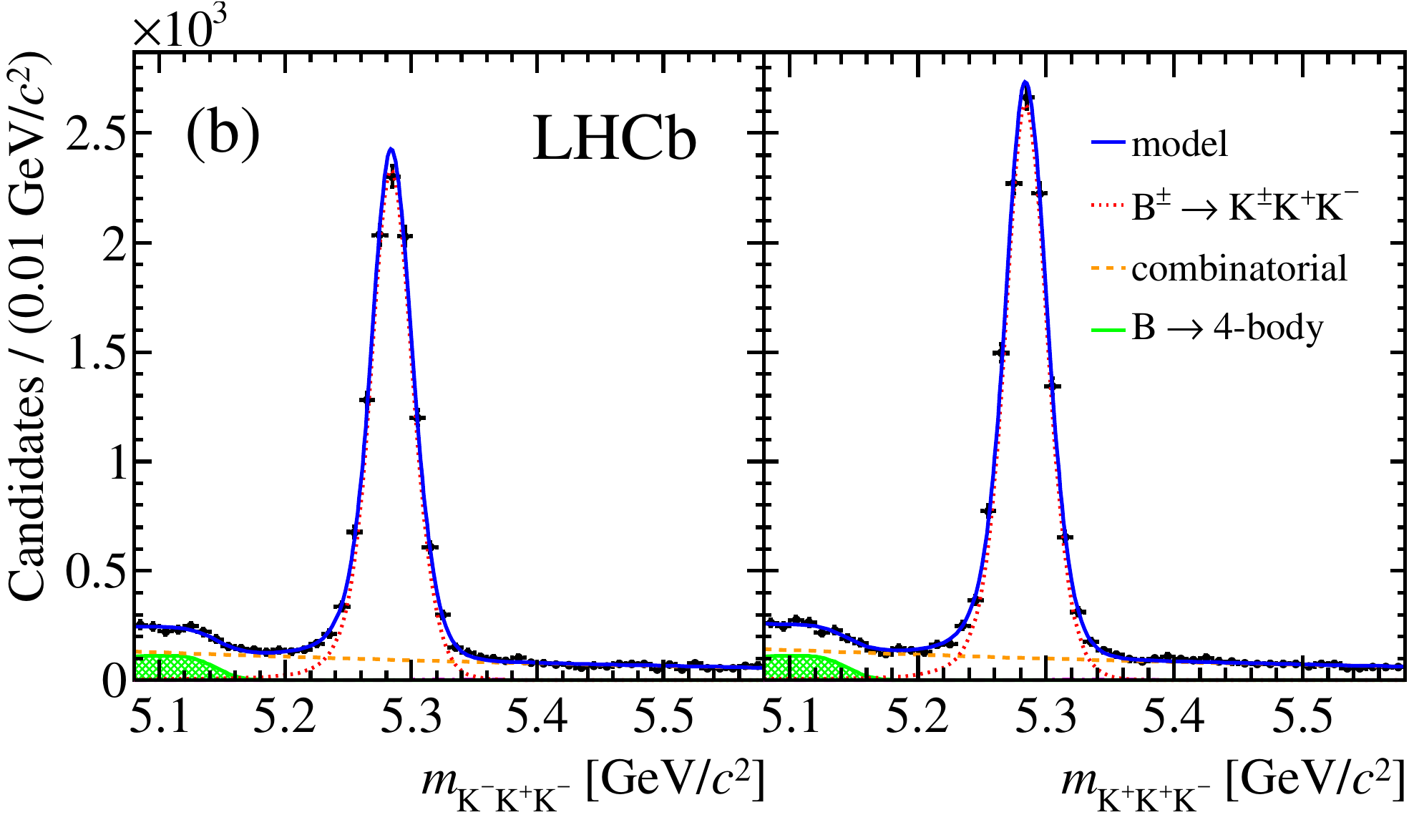}
\caption{Distributions of the \B-candidate invariant mass for (a) \BuToKpipi
decays and (b) \BuToKKK decays.  The left (right) plots show the \Bm (\Bp)
decays.}
\label{fig:Khh-mass-fits}
\end{figure}

The results of the fit to the data sample are shown in
Figure~\ref{fig:Khh-mass-fits} and the values of the \CP asymmetries are found
to be
\begin{eqnarray}
\nonumber
\ACP(\BuToKpipi) &=& \phantom{-}0.032 \pm 0.008 \stat \pm 0.004 \syst \pm 0.007 (\jpsi\Kp) \,, \\
\nonumber
\ACP(\BuToKKK)   &=& -0.043 \pm 0.009 \stat \pm 0.003 \syst \pm 0.007 (\jpsi\Kp) \,.
\end{eqnarray}
The significance of \CP violation in each decay mode is $2.8\,\sigma$ and
$3.7\,\sigma$, respectively.

\begin{figure}[htb]
\centering
\includegraphics[width=0.49\textwidth]{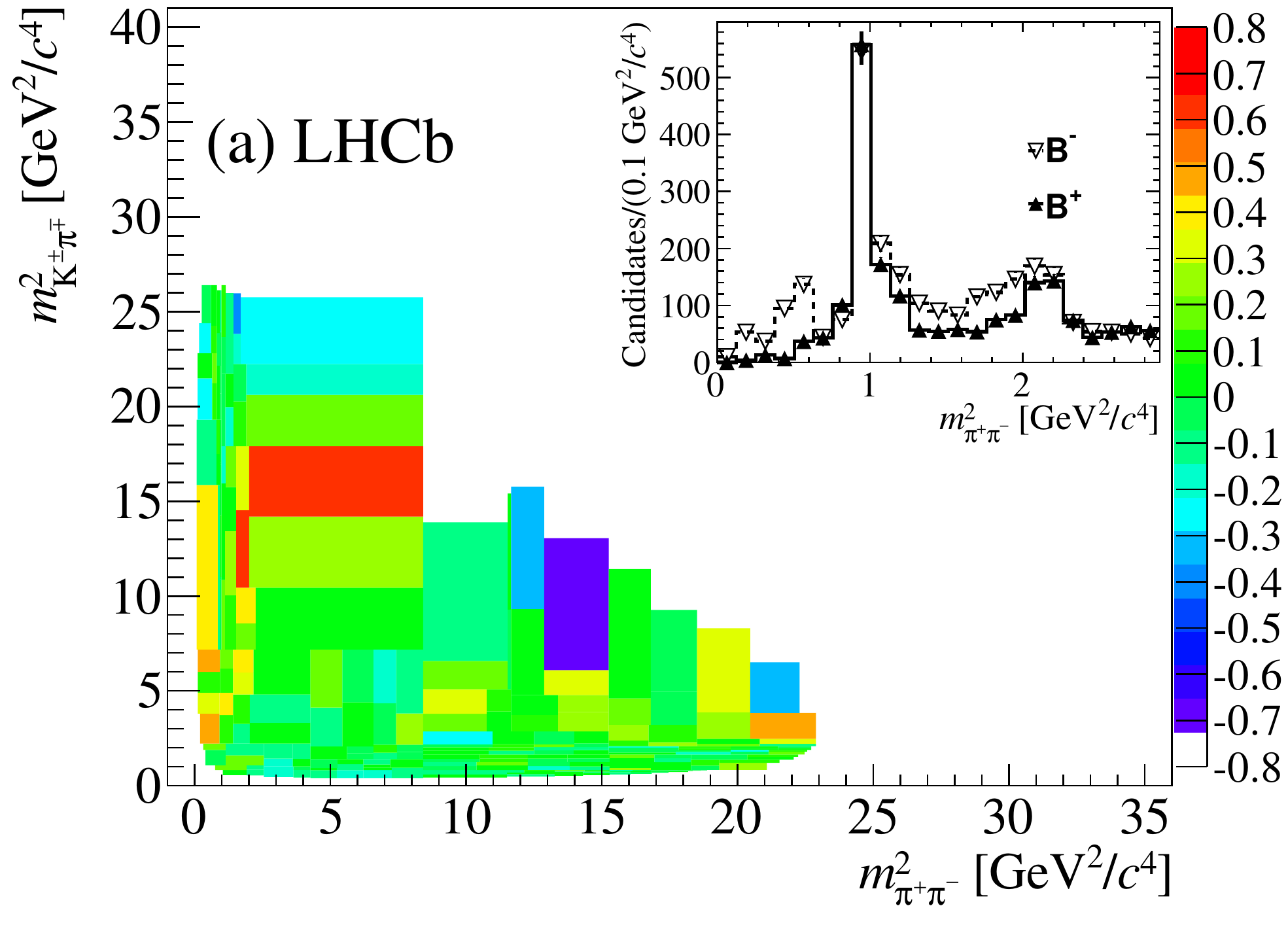}
\includegraphics[width=0.49\textwidth]{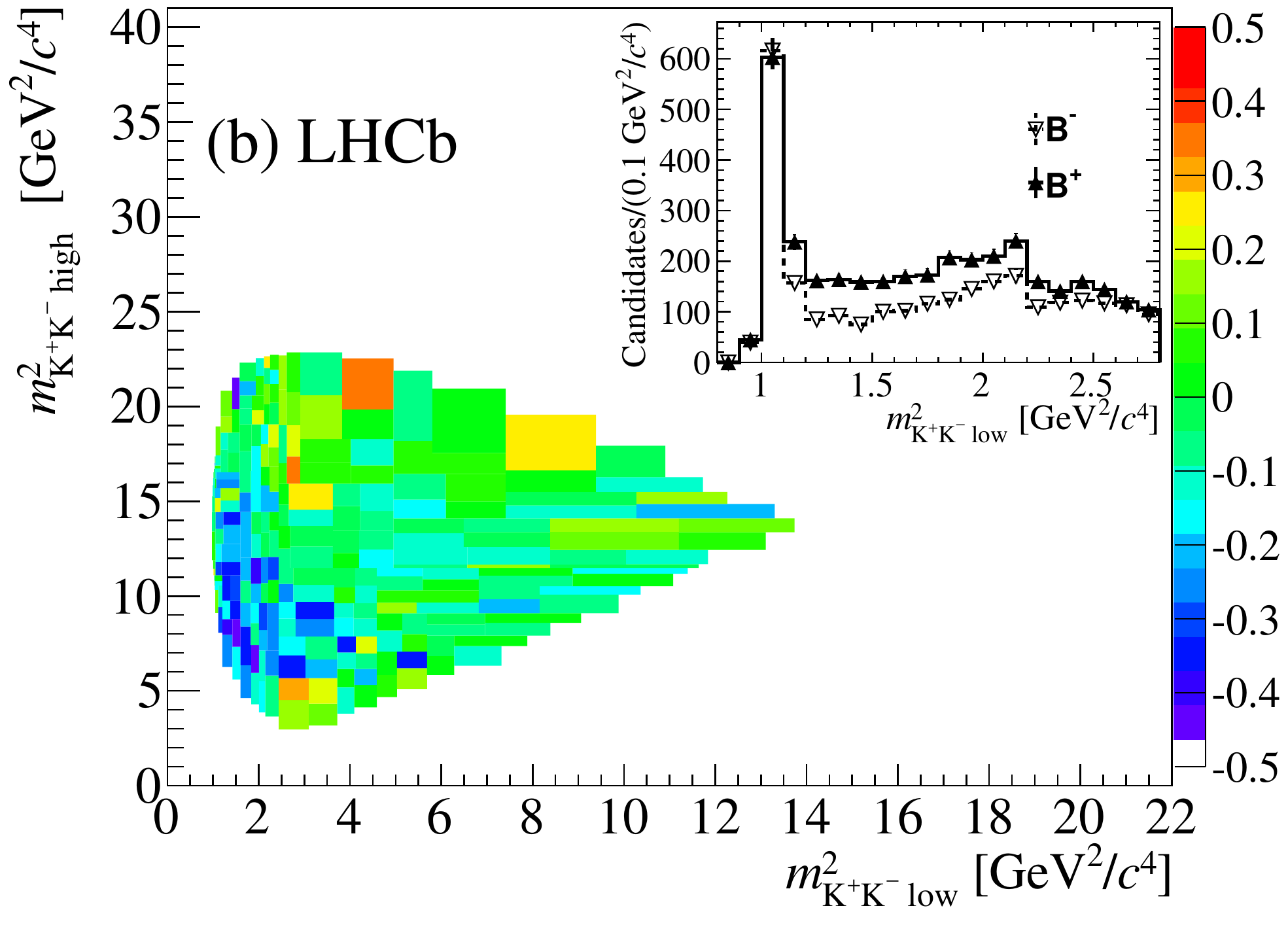}
\caption{Variation of the raw asymmetry over the Dalitz plot in (a) \BuToKpipi
and (b) \BuToKKK decays.}
\label{fig:Khh-DP}
\end{figure}

The variation of the raw asymmetry over the Dalitz plot is also studied and the
results are shown in Figure~\ref{fig:Khh-DP}.  In some regions there are extremely
large asymmetries present, in particular around the \rhoz resonance in \BuToKpipi
but also in regions that are not clearly associated with a resonance.  The local
\CP asymmetries in the region where $m^2_{\Kp\pim} < 15 \,(\!\gevcc)^2$ and
$0.08 < m^2_{\pipi} < 0.66 \,(\!\gevcc)^2$ in \BuToKpipi and in the region
$m^2_{\KpKm {\rm high}} < 15 \,(\!\gevcc)^2$ and
$1.2 < m^2_{\KpKm {\rm low}} < 2.0 \,(\!\gevcc)^2$
are determined to be
\begin{eqnarray}
\nonumber
\ACP_{\rm local}(\BuToKpipi) &=& \phantom{-}0.678 \pm 0.078 \stat \pm 0.032 \syst \pm 0.007 (\jpsi\Kp) \,, \\
\nonumber
\ACP_{\rm local}(\BuToKKK)   &=& -0.226 \pm 0.020 \stat \pm 0.004 \syst \pm 0.007 (\jpsi\Kp) \,,
\end{eqnarray}
respectively.

\begin{figure}[htb]
\centering
\includegraphics[width=0.49\textwidth]{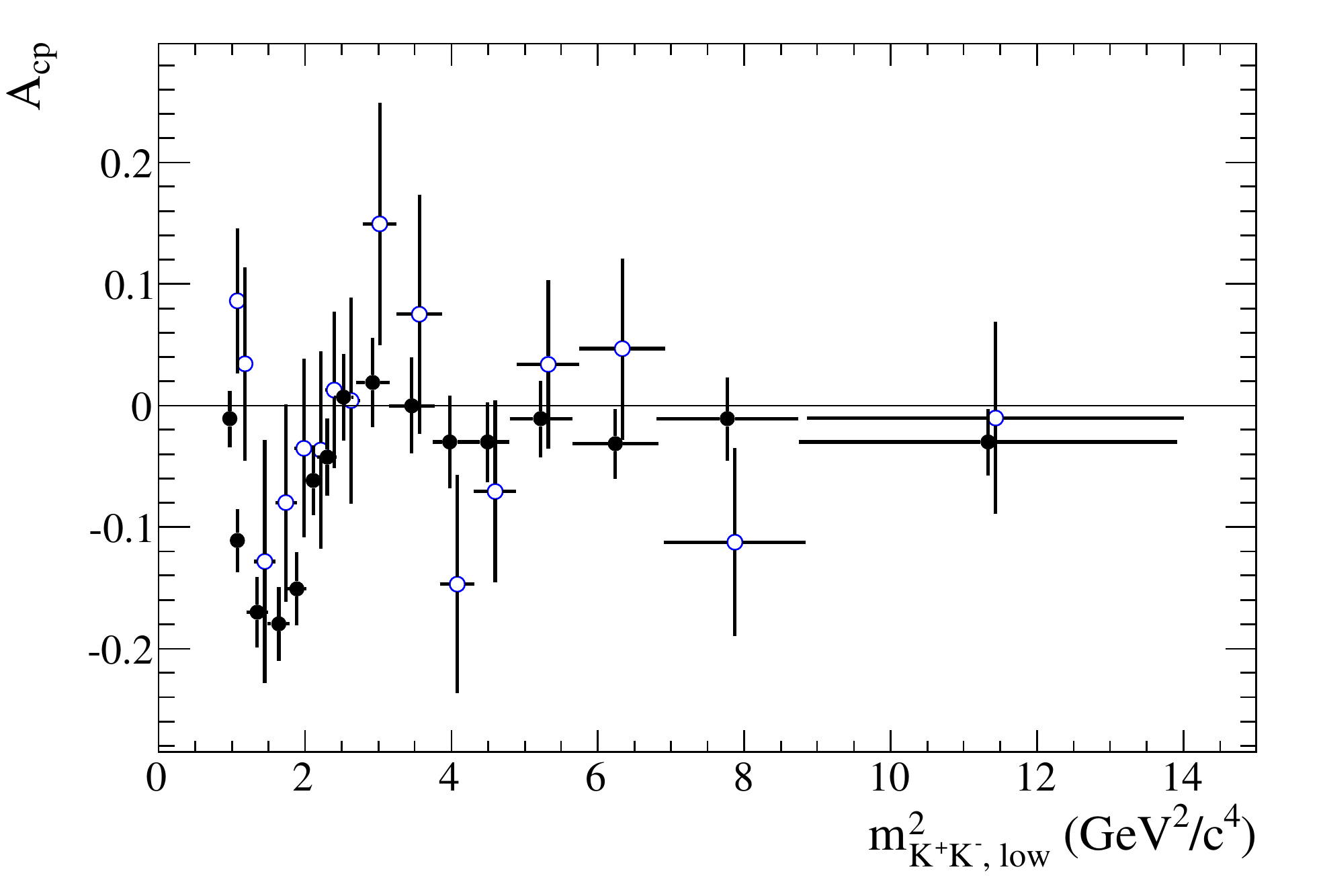}
\includegraphics[width=0.49\textwidth]{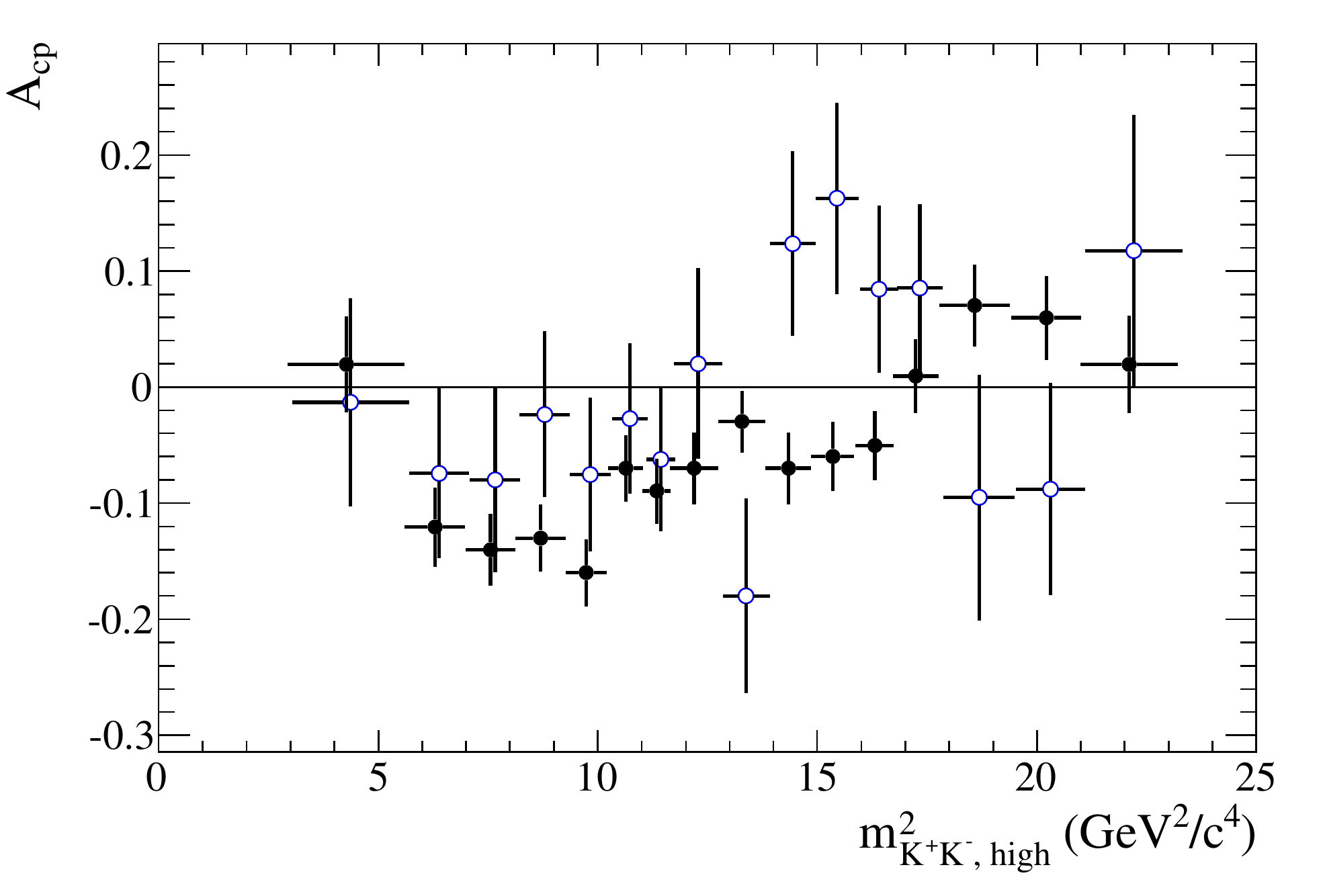}
\caption{Asymmetry as a function of (left) $m_{\KpKm {\rm low}}^2$
(right) $m_{\KpKm {\rm high}}^2$ for \BuToKKK decays.
The \babar (LHCb) data are the open (filled) circles.}
\label{fig:KKK-asym-comp}
\end{figure}

The \babar experiment has recently made an update of their analysis of \BuToKKK,
in order to provide a direct comparison with the LHCb results for the asymmetry
as a function of $m_{\KpKm}$~\cite{Lees:2013ngt}.
This comparison is shown in Figure~\ref{fig:KKK-asym-comp}.
The shapes of the distributions are extremely similar, albeit with a small
offset, which is determined to be $0.045 \pm 0.021$ ($0.053 \pm 0.021$) for the
$m_{\KpKm {\rm low}}^2$ ($m_{\KpKm {\rm high}}^2$) spectrum.
However, it must be remembered that the LHCb distribution is that of the raw
asymmetry and hence has not been corrected for production and detection effects.
These are of the order of 1\% and act in the direction to decrease the mild
discrepancy.

Very similar findings to those from \BuToKhh decays are made in a preliminary
analysis of \BuTopipipi and \BuToKKpi~\cite{LHCb-CONF-2012-028}, both in terms
of the large local asymmetries and the opposite sign of the asymmetries between
the two modes.  In addition, the local asymmetries are observed mainly in regions
not clearly associated with a well established resonance.
This could indicate that \decay{\pipi}{\KpKm} rescattering is playing a role
in the generation of the strong phase difference.
Amplitude analyses of these modes using the larger dataset now available at LHCb
(3\invfb) will provide more information to resolve this puzzle.

\section{\boldmath Dynamics of \BuTopph decays}

The large asymmetries seen in \BuTohhh decays raise the question about the role
of $\pipi \leftrightarrow \KpKm$ rescattering in these modes.
The closely related decays \BuTopph can shed some light on this issue since it
is expected that $\Ph^+\Ph^- \leftrightarrow \ppbar$ rescattering should be much
smaller.
The threshold enhancements seen in many \decay{\B}{\ppbar X} decays
provide further motivation for studying these decays.
The analysis, which uses the LHCb 1.0\invfb data sample collected during 2011,
studies the dynamics of the decays as well as the \CP asymmetries.
Full details can be found in Ref.~\cite{Aaij:2013fla}.

\begin{figure}[htb]
\centering
\includegraphics[width=0.49\textwidth]{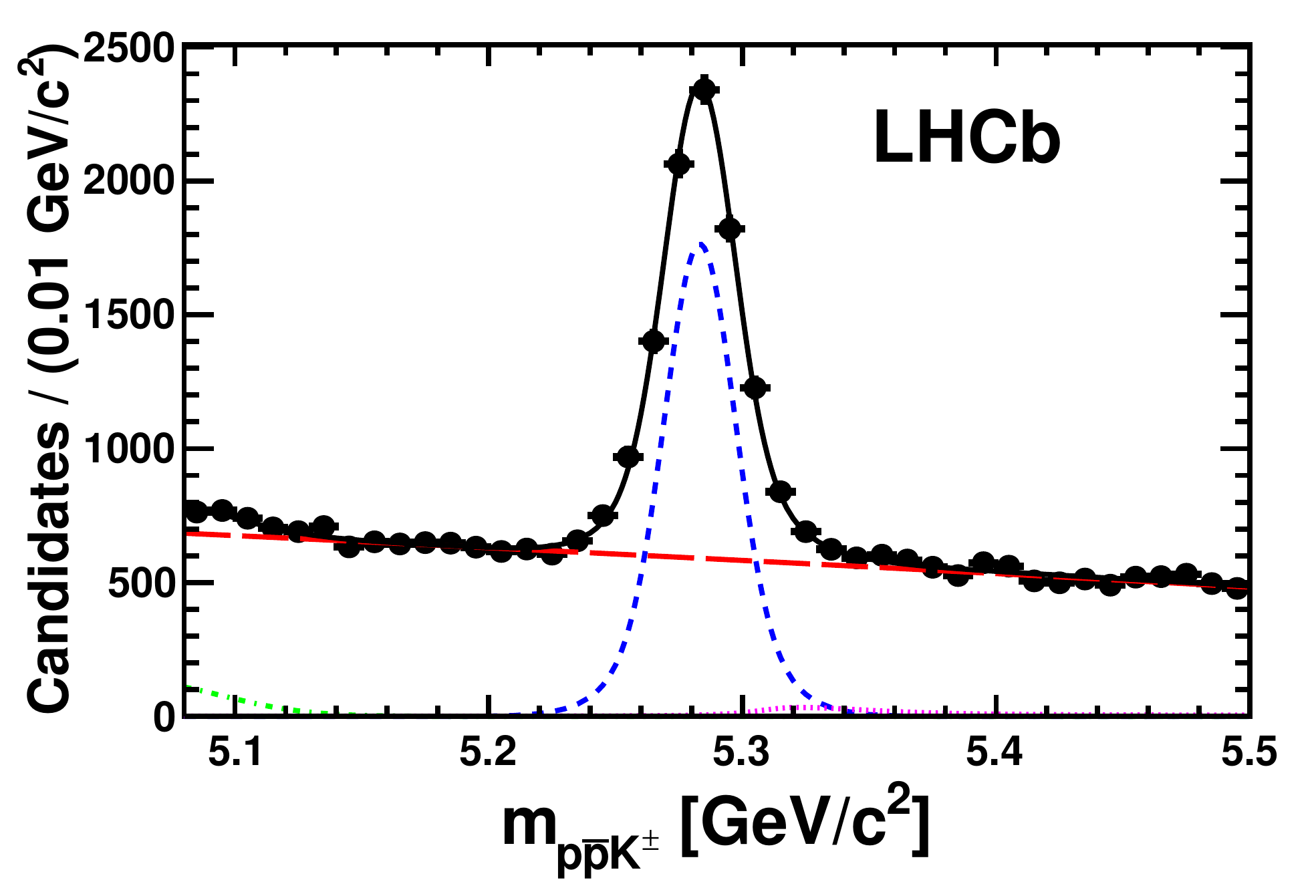}
\includegraphics[width=0.49\textwidth]{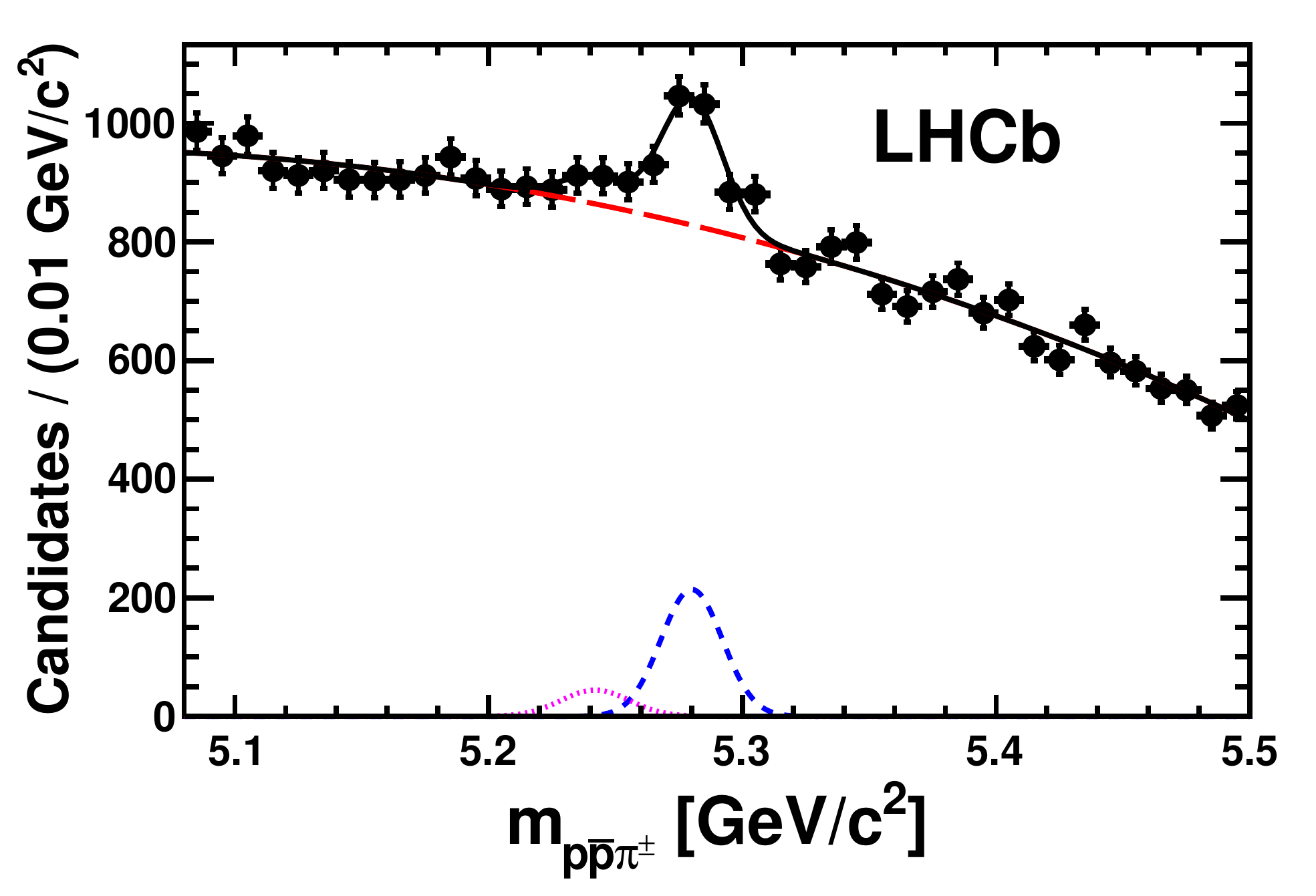}
\caption{Distributions of the \B-candidate invariant mass for (left) \BuToppK
decays and (right) \BuTopppi decays.}
\label{fig:pph-mass-fits}
\end{figure}

\begin{figure}[htb]
\centering
\includegraphics[width=0.49\textwidth]{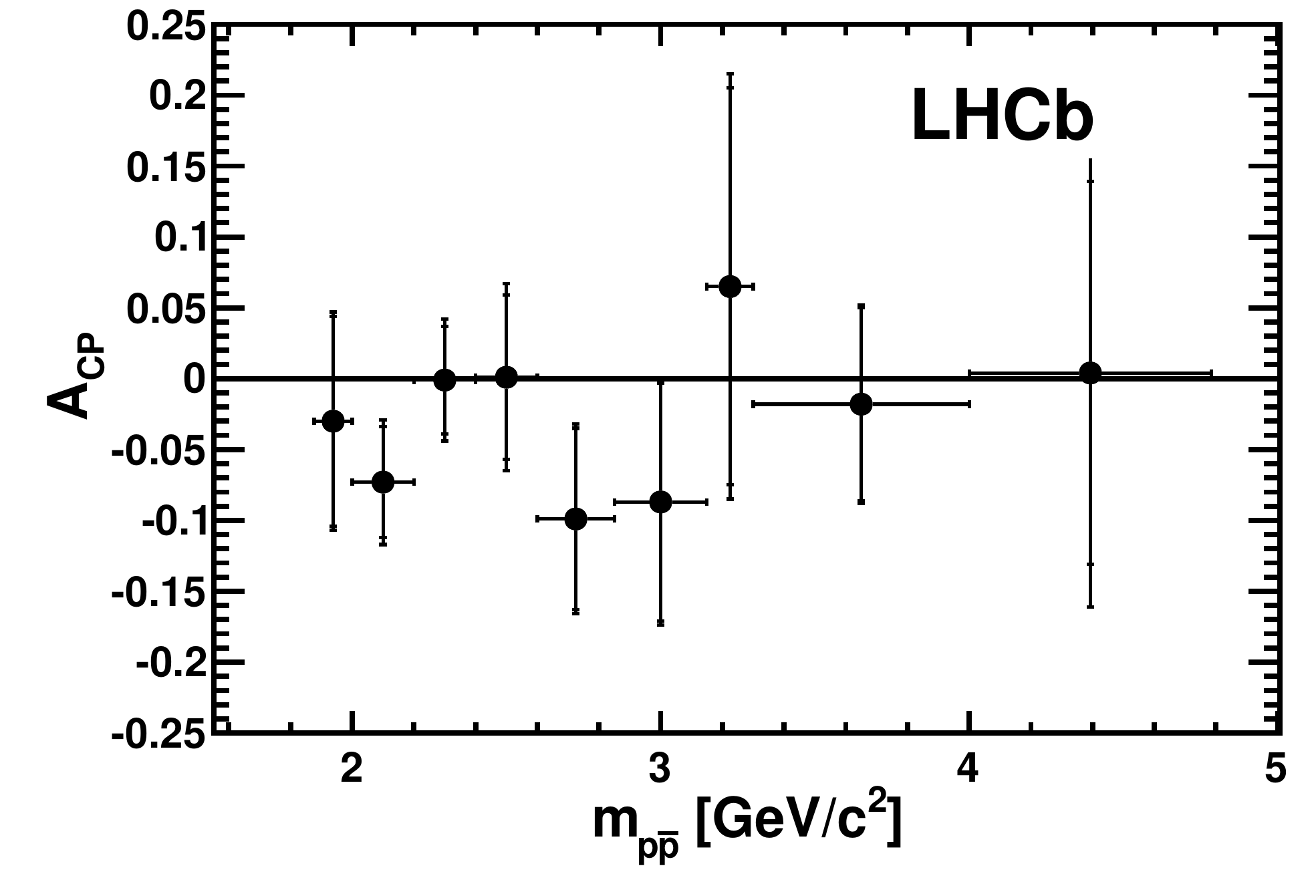}
\includegraphics[width=0.49\textwidth]{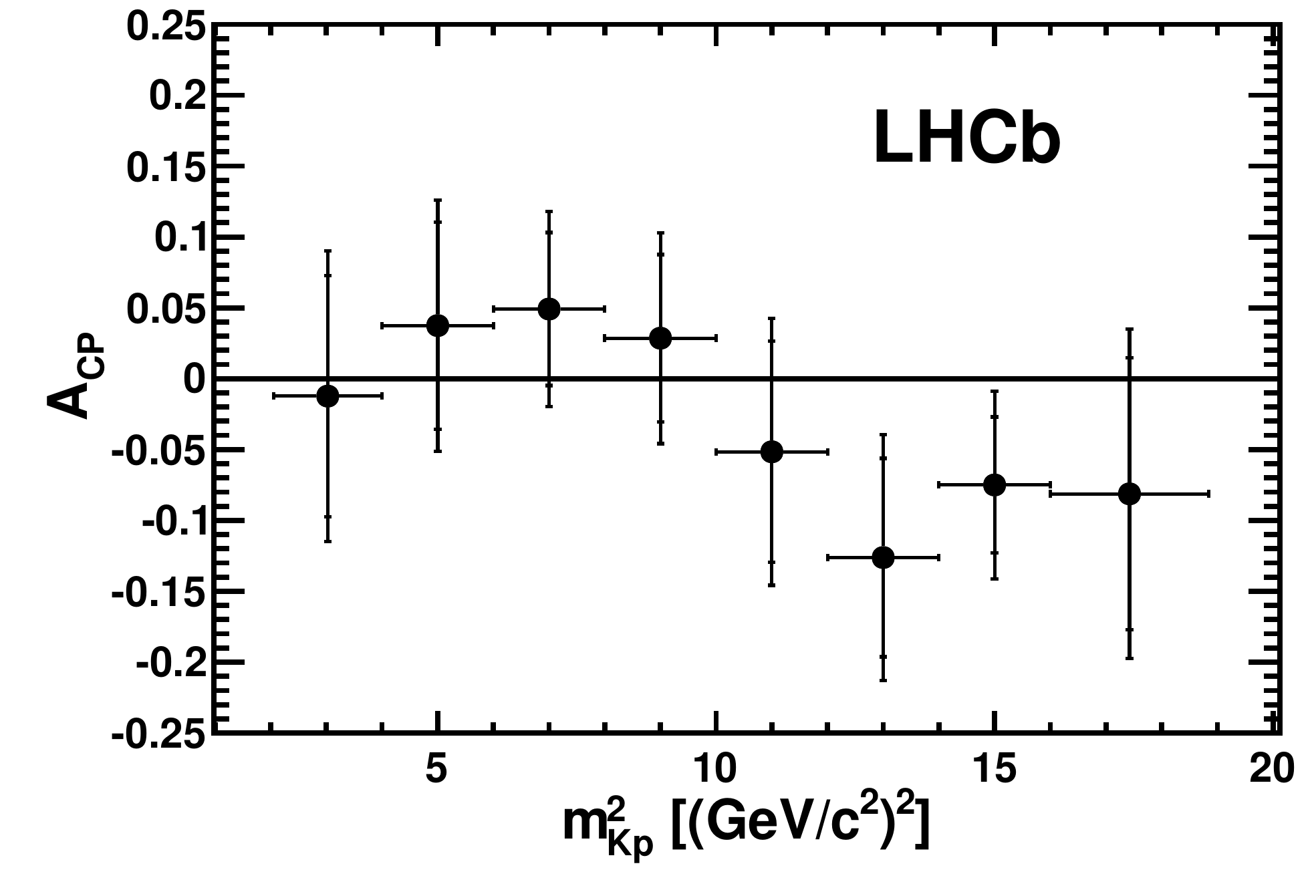}
\caption{\CP asymmetry as a function of (left) $m_{\ppbar}$
(right) $m_{\Kproton}$ for \BuToppK decays.}
\label{fig:ppK-ACP}
\end{figure}

Fits to the \B-candidate invariant mass distribution, shown in
Figure~\ref{fig:pph-mass-fits}, yield $7029 \pm 139$ ($656 \pm 70$) signal events
for the mode \BuToppK (\BuTopppi), where the uncertainties are statistical only.
The fit model contains contributions from signal, cross-feed (where the kaon in
the signal mode is mis-identified as a pion or vice versa), combinatorial and
partially-reconstructed backgrounds.
The \CP asymmetries for \BuToppK are determined  by repeating the fits to the
\B-candidate invariant mass in bins of both the \ppbar and \Kproton invariant
masses and separating by the charge of the \B candidate.  The results are shown
in Figure~\ref{fig:ppK-ACP} and are consistent with zero in all bins, albeit
with large uncertainties.

\begin{figure}[htb]
\centering
\includegraphics[width=0.49\textwidth]{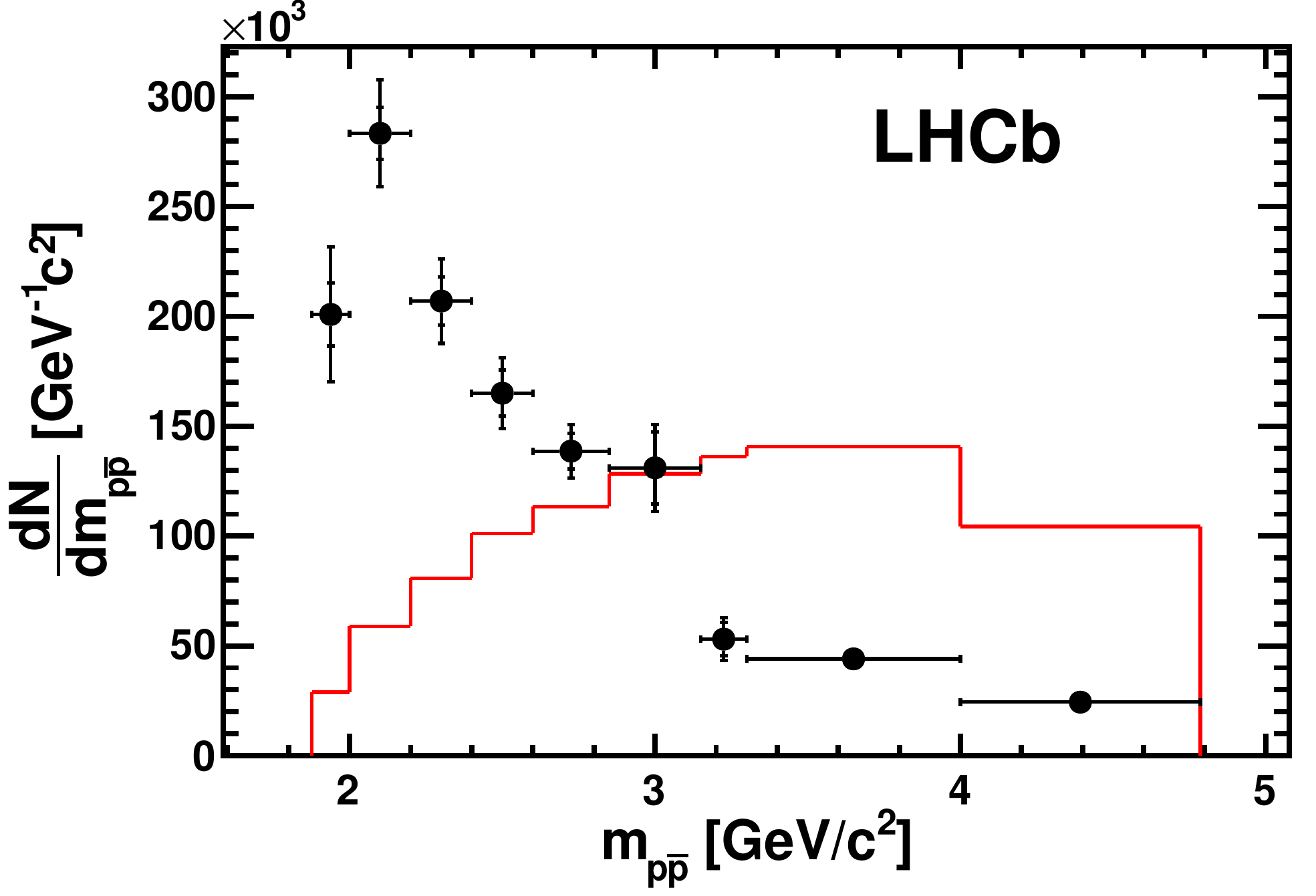}
\includegraphics[width=0.49\textwidth]{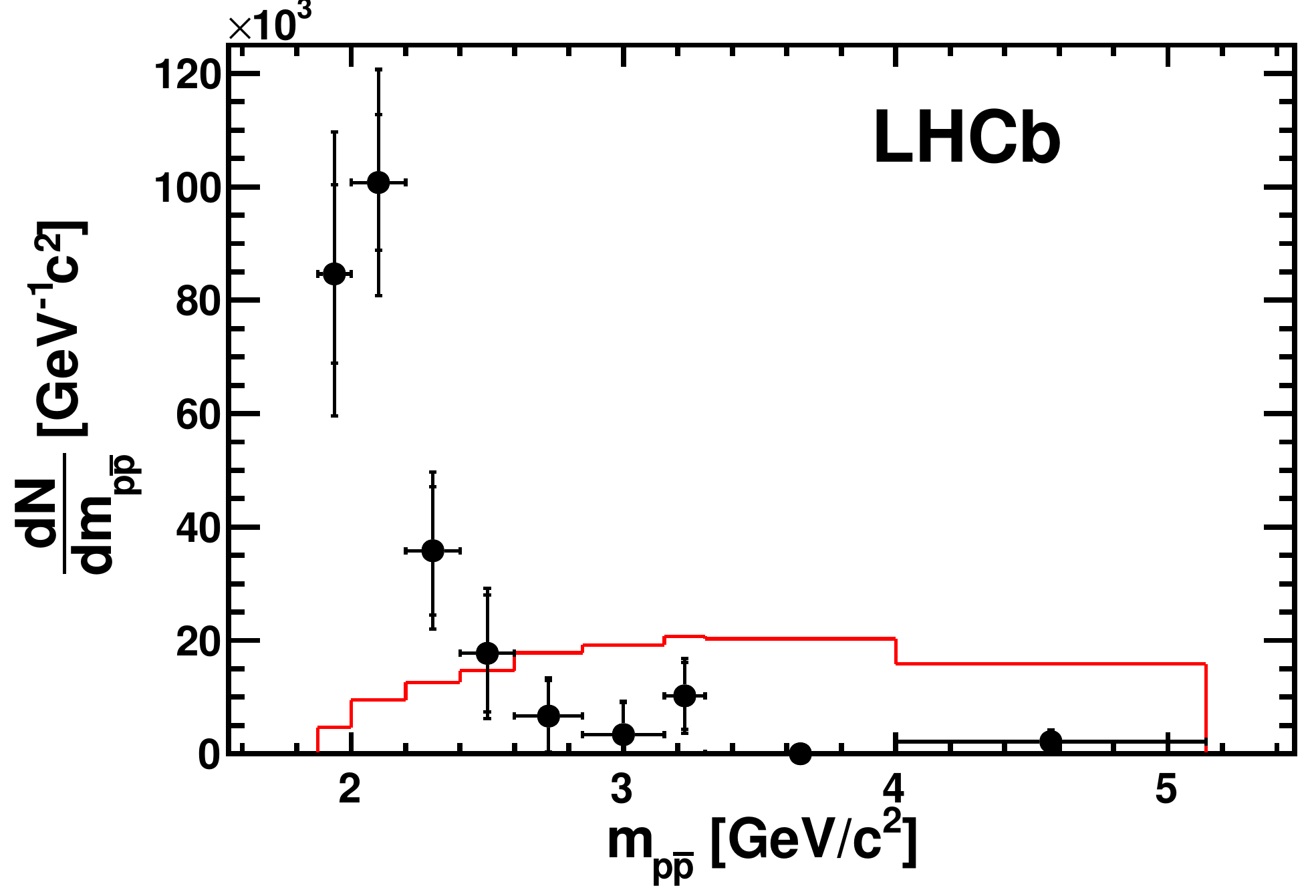}
\caption{Differential production spectra as a function of $m_{\ppbar}$ for
(left) \BuToppK decays (right) \BuTopppi decays.}
\label{fig:pph-invmass}
\end{figure}

\begin{figure}[htb]
\centering
\includegraphics[width=0.49\textwidth]{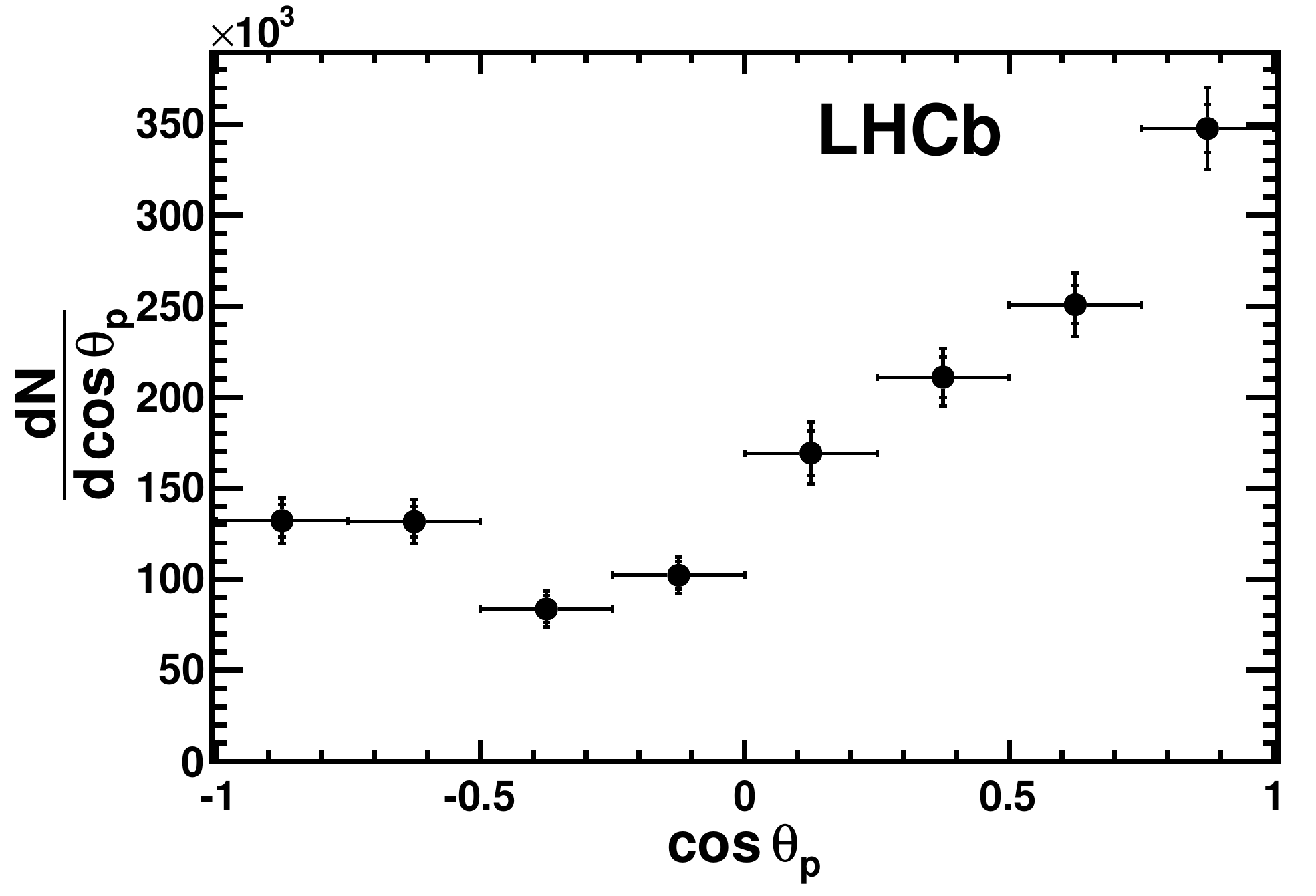}
\includegraphics[width=0.49\textwidth]{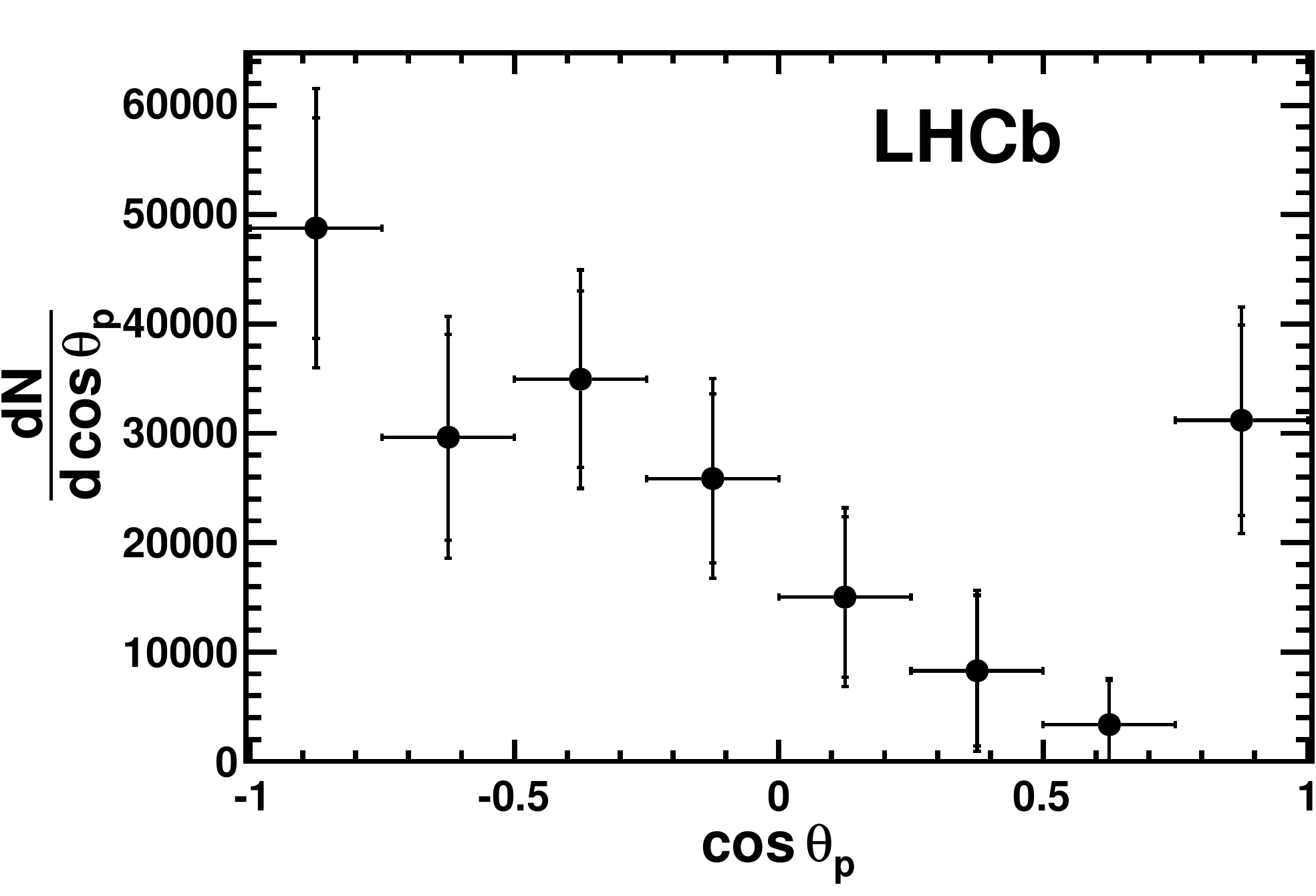}
\caption{Differential production spectra as a function of $\cos\theta_\proton$ for
(left) \BuToppK decays (right) \BuTopppi decays.}
\label{fig:pph-costheta}
\end{figure}

\begin{figure}[htb]
\centering
\includegraphics[width=0.49\textwidth]{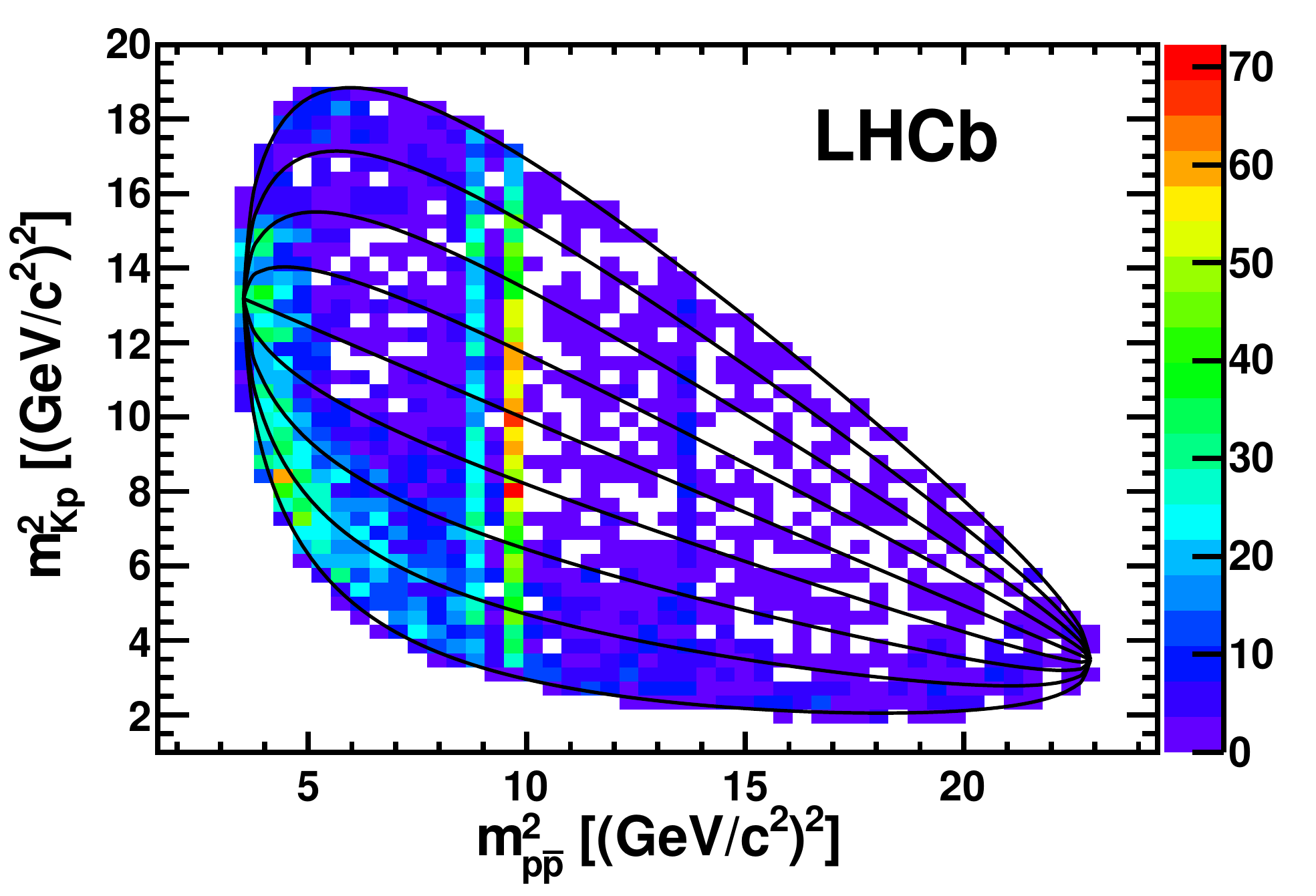}
\caption{Dalitz plot distribution for \BuToppK signal events.
The black solid curves are lines of constant $\cos\theta_\proton$.}
\label{fig:ppK-DP}
\end{figure}

\begin{figure}[htb]
\centering
\includegraphics[width=0.49\textwidth]{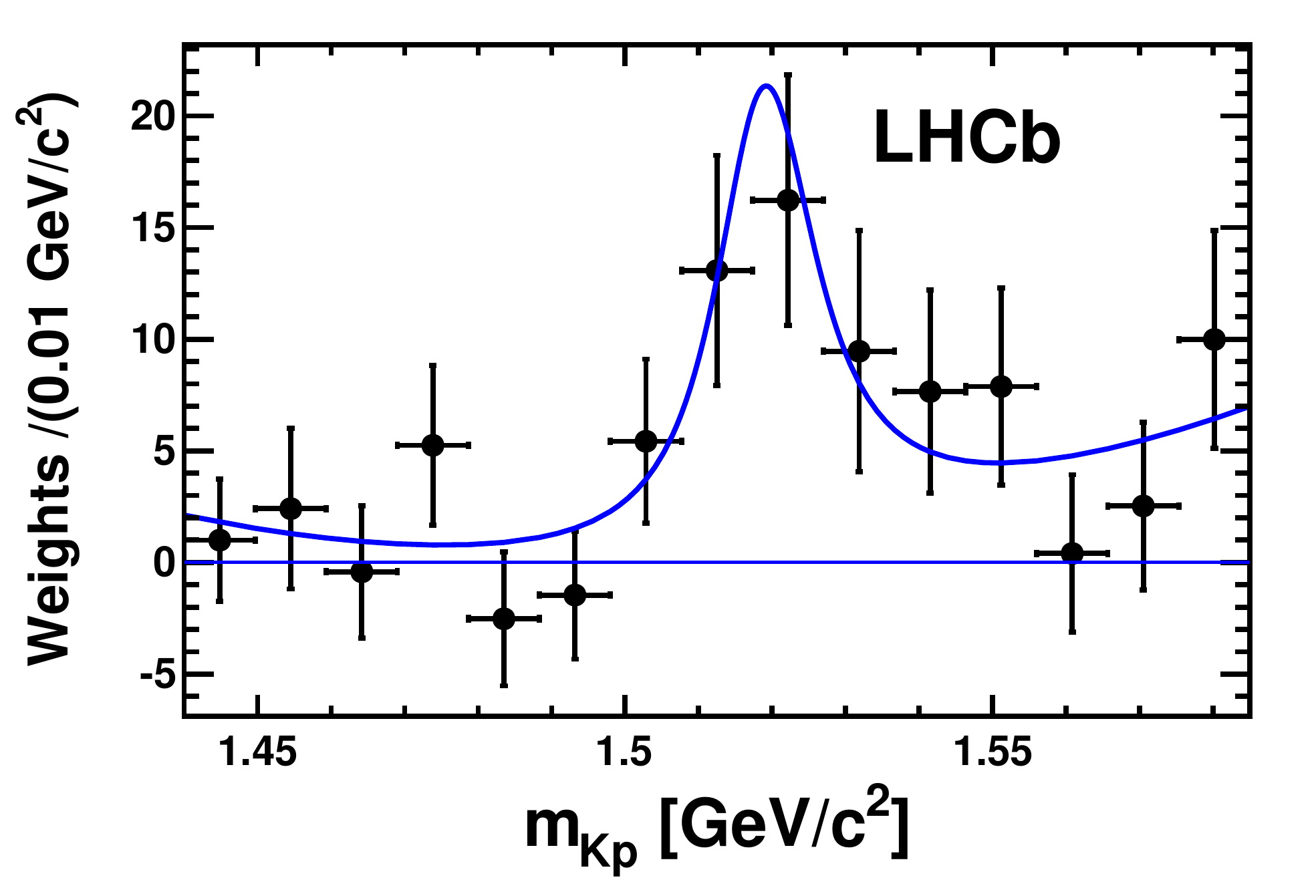}
\caption{Distribution of \Kproton invariant mass for \BuToppK signal events
in the region $1.440 < m_{\Kproton} < 1.585 \gevcc$.}
\label{fig:ppK-mKp}
\end{figure}

The decay dynamics are studied by constructing differential production spectra
as a function of the invariant masses and the cosine of the angle,
$\theta_\proton$, between the daughter meson and the opposite-sign baryon in
the \ppbar rest frame.
The distributions as a function of \ppbar invariant mass are shown in
Figure~\ref{fig:pph-invmass} and show very clear threshold enhancement behaviour,
similar to other \decay{\B}{\ppbar X} decays.
The distributions as a function of $\cos\theta_\proton$ are shown in
Figure~\ref{fig:pph-costheta} and exhibit strikingly opposite behaviour between
the two decay modes, the forward/backward asymmetries being
\begin{eqnarray}
\nonumber
\AFB(\BuToppK) &=& \phantom{-}0.370 \pm 0.018 \stat \pm 0.016 \syst \,, \\
\nonumber
\AFB(\BuTopppi) &=& -0.392 \pm 0.117 \stat \pm 0.015 \syst \,.
\end{eqnarray}
This behaviour can also clearly be seen when examining the \BuToppK Dalitz plot
shown in Figure~\ref{fig:ppK-DP}, which has been background-subtracted using
the \sPlot technique~\cite{Pivk:2004ty}.
The other clear features are the vertical bands at high \ppbar invariamt mass,
which are contributions from charmonium intermediate states.  These have been
studied separately in an analysis reported in Ref.~\cite{Aaij:2013rha}.
There is also some structure at low $m_{\Kproton}$, which is shown more clearly
in the signal \sPlot invariant mass projection in Figure~\ref{fig:ppK-mKp}.
A two-dimensional fit to the \B-candidate invariant mass and $m_{\Kproton}$ is
performed in this region in order to extract the yield of the $\Lbar(1520)$
resonance.  The signal is found to have a significance of $5.1\,\sigma$, which
constitutes first observation of the decay \BuToLbp with a branching fraction of
\begin{eqnarray}
\nonumber
\BF(\BuToLbp) = (3.9 \,^{+1.0}_{-0.9} \stat \pm 0.1 \syst \pm 0.3 ({\rm BF})) \times 10^{-7} \,,
\end{eqnarray}
where the third uncertainty is from the branching fraction of \BuToJPsiK,
\JPsiTopp.

\clearpage

\section{\boldmath Results from \BdTohhpiz decays}

\begin{figure}[htb]
\centering
\includegraphics[width=0.49\textwidth]{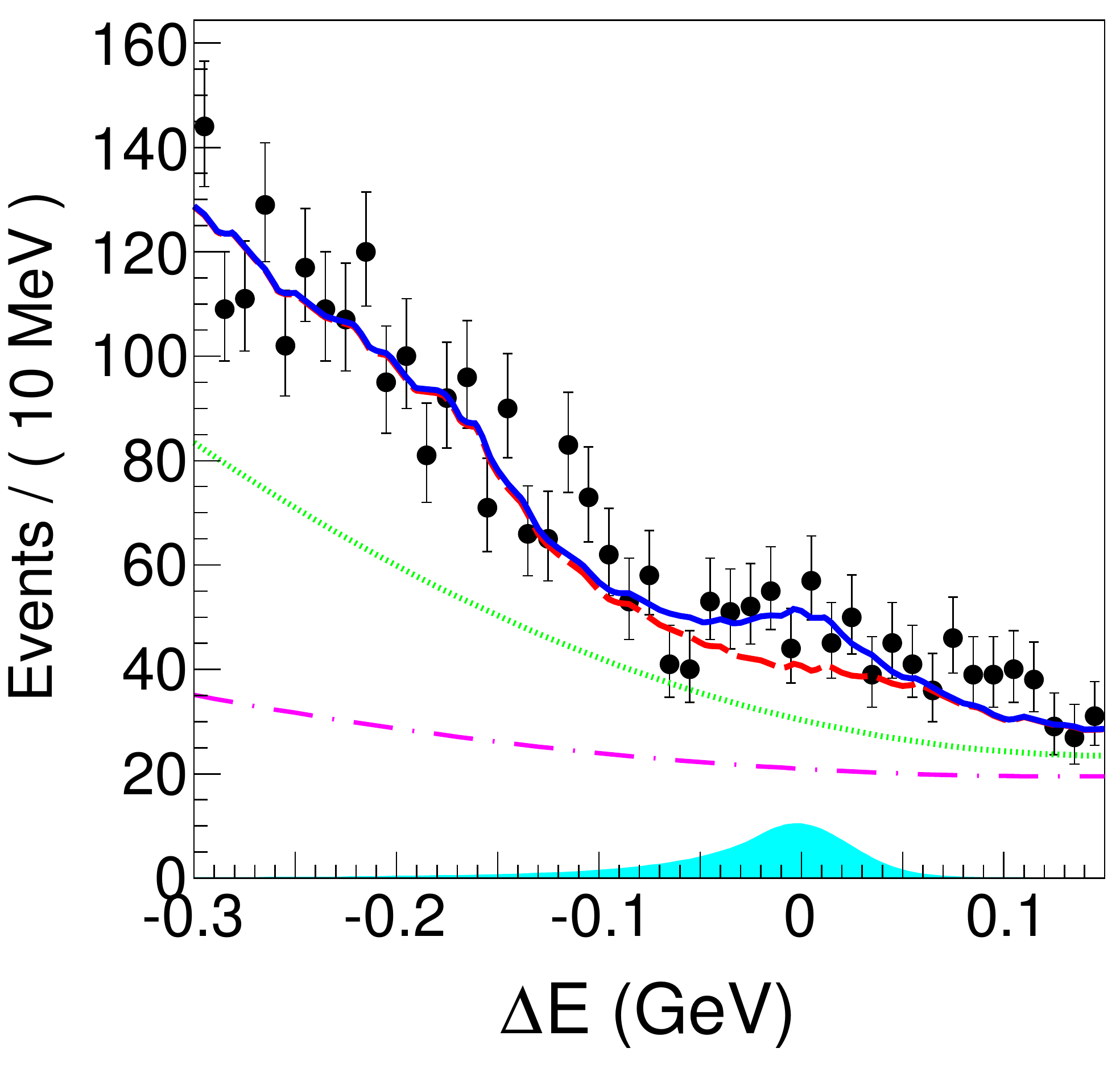}
\includegraphics[width=0.49\textwidth]{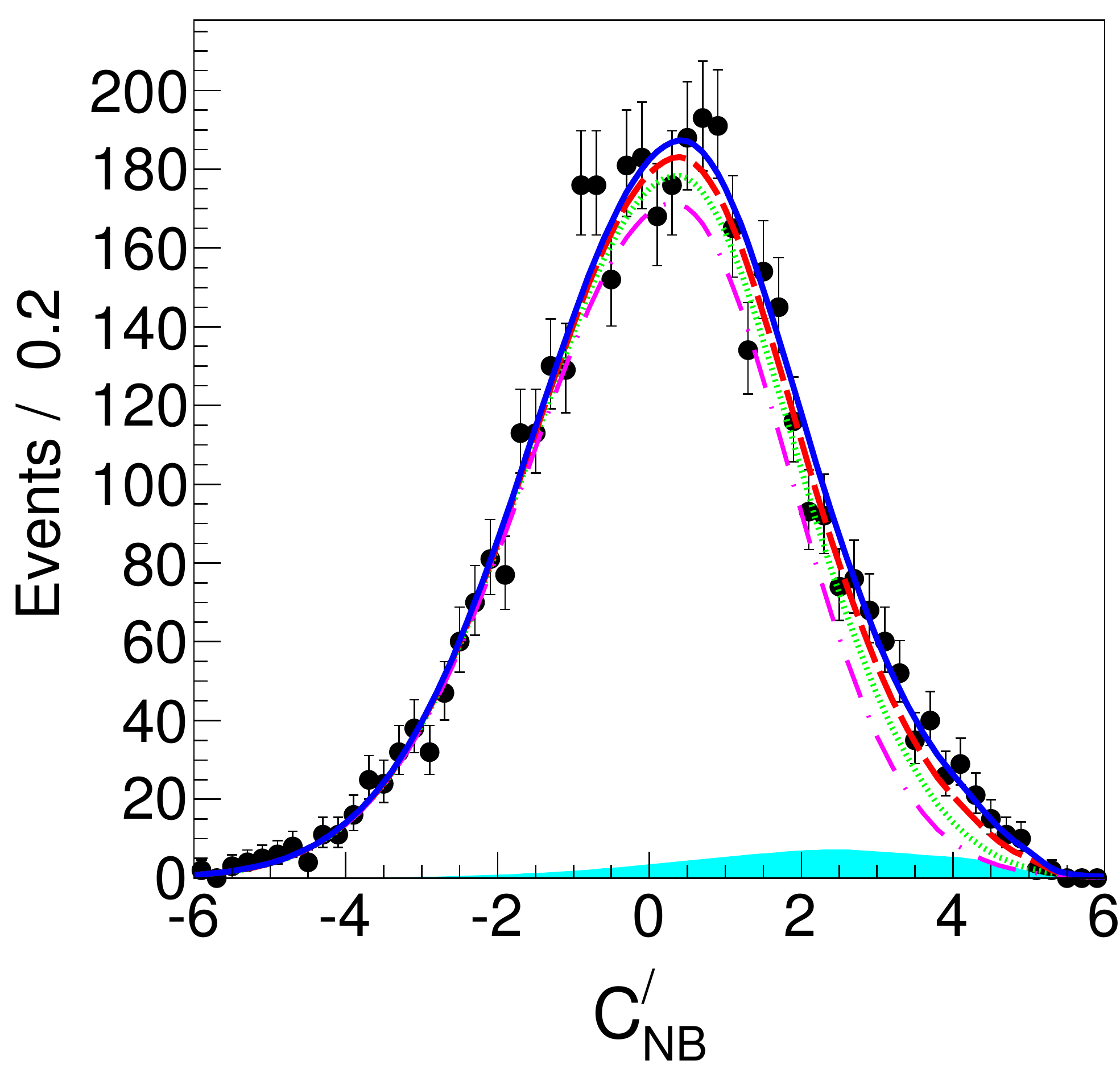}
\caption{Projections of the maximum likelihood fit to \BdToKKpiz candidate
events for the variables (left) $\Delta E$ and (right) the neural network of
event-shape variables.}
\label{fig:KKpiz-fit}
\end{figure}

The Belle collaboration have recently reported the results of a search for the
decay \BdToKKpiz, using a data sample of 772 million \BBbar pairs.
Full details of the analysis are given in Ref.~\cite{Gaur:2013uou}.
A fit is performed to $\Delta E$, the difference between the energy of the \B
candidate and the beam energy, and the output of a neural network of event-shape
variables.  The latter variable is a powerful discriminant against the dominant
background from continuum light-quark production.
The fit yields $299 \pm 83$ signal events, where the uncertainty is statistical
only.  The projections of the fit are shown in Figure~\ref{fig:KKpiz-fit}.
The signal has a significance of $3.5\,\sigma$, which constitutes the first
evidence of this decay, with a branching fraction of
\begin{eqnarray}
\BF(\BdToKKpiz) = ( 2.17 \pm 0.60 \stat \pm 0.24 \syst ) \times 10^{-6} \,.
\end{eqnarray}

\begin{figure}[htb]
\centering
\includegraphics[width=0.49\textwidth]{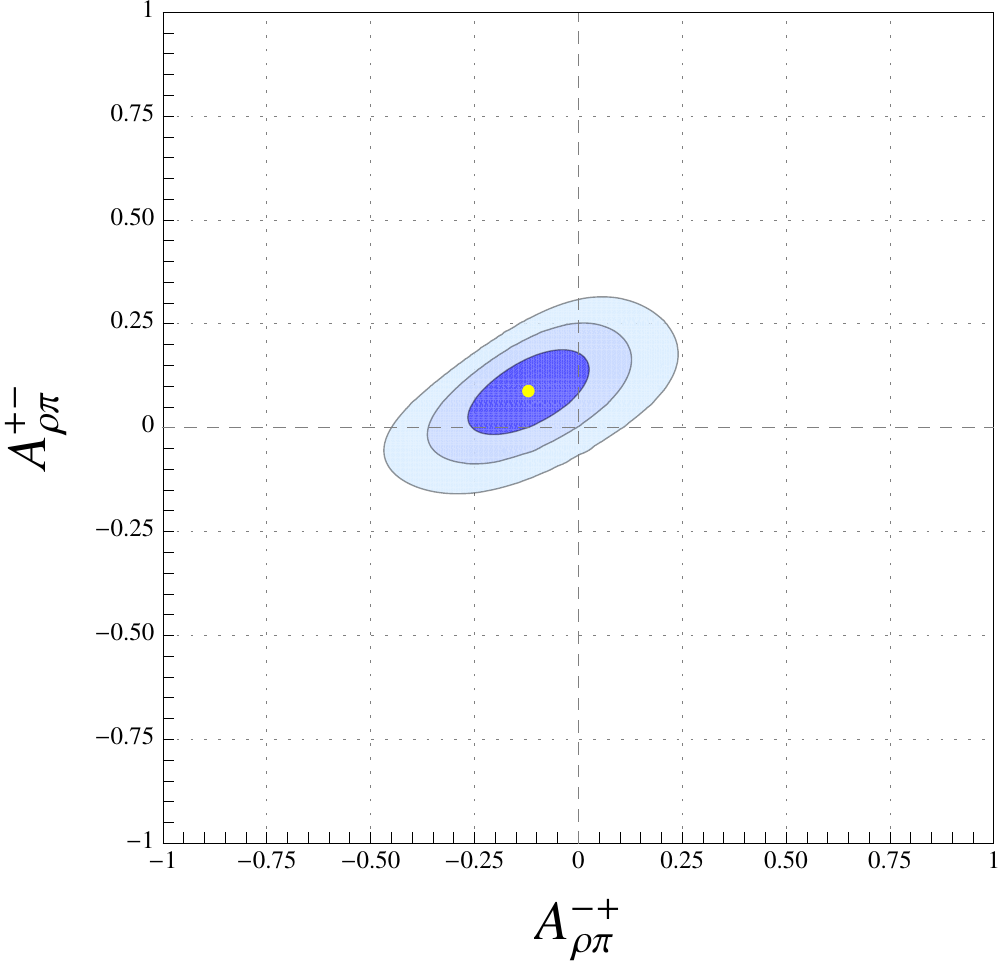}
\caption{Likelihood scan in the $A_{\Prho\Ppi}^{+-}$ vs. $A_{\Prho\Ppi}^{-+}$
plane.}
\label{fig:rhopi-ACP}
\end{figure}

The \babar collaboration have recently updated their time-dependent Dalitz-plot
analysis of the decay \BdTopipipiz to use their full \FourS dataset of
471 million \BBbar pairs.
The primary goal of this analysis is to measure the CKM angle $\alpha$ using the
Snyder-Quinn method~\cite{Snyder:1993mx}.
Full details of the analysis can be found in Ref.~\cite{Lees:2013nwa}.
A thorough robustness study was conducted, which showed that while the extraction
of the fit parameters and most of the derived quasi-two-body parameters was
robust, unfortunately the extraction of $\alpha$ itself was not.
However, hints of direct \CP violation were seen in the two parameters
\begin{eqnarray}
A_{\Prho\Ppi}^{+-} &=& \frac{\Gamma(\BdbTorhompip) - \Gamma(\BdTorhoppim)}
                            {\Gamma(\BdbTorhompip) + \Gamma(\BdTorhoppim)} \,, \\
A_{\Prho\Ppi}^{-+} &=& \frac{\Gamma(\BdbTorhoppim) - \Gamma(\BdTorhompip)}
                            {\Gamma(\BdbTorhoppim) + \Gamma(\BdTorhompip)} \,.
\end{eqnarray}
The result of the 2D scan for these parameters is shown in
Figure~\ref{fig:rhopi-ACP}.  The consistentcy with the no direct \CP
violation point is quantified as $\Delta\chisq = 6.42$.

\section{\boldmath Studies of \BdsToKShh decays}

Time-dependent flavour-tagged Dalitz-plot analyses of \B decays to \KShh final
states are sensitive to mixing-induced \CP-violating phases.  For example, the
recent \babar measurement
$\beta_{\rm eff}(\Pphi\KS) = (21 \pm 6 \pm 2)^{\degrees}$
in the decay \BdToKSKK~\cite{Lees:2012kxa}.
Such an analysis is not possible with the current LHCb statistics, however it
is possible to search for the previously unobserved \Bs decays to these final
states.

The analysis, which uses the LHCb 1.0\invfb data sample collected during 2011, has
separate optimisations of the selection for the suppressed and favoured decays
in each final state.  In addition, most of the reconstructed \KS mesons decay
downstream of the LHCb Vertex Locator and so do not have information from that
subdetector, while the remaining $\sim \frac{1}{3}$ do have such information.
This leads to rather different efficiencies for the two types of \KS candidates
(referred to as Downstream and Long, respectively)
and hence the need to treat each category separately in the analysis.
Full details of the analysis can be found in Ref.~\cite{Aaij:2013uta}.

\begin{figure}[htb]
\centering
\includegraphics[width=0.49\textwidth]{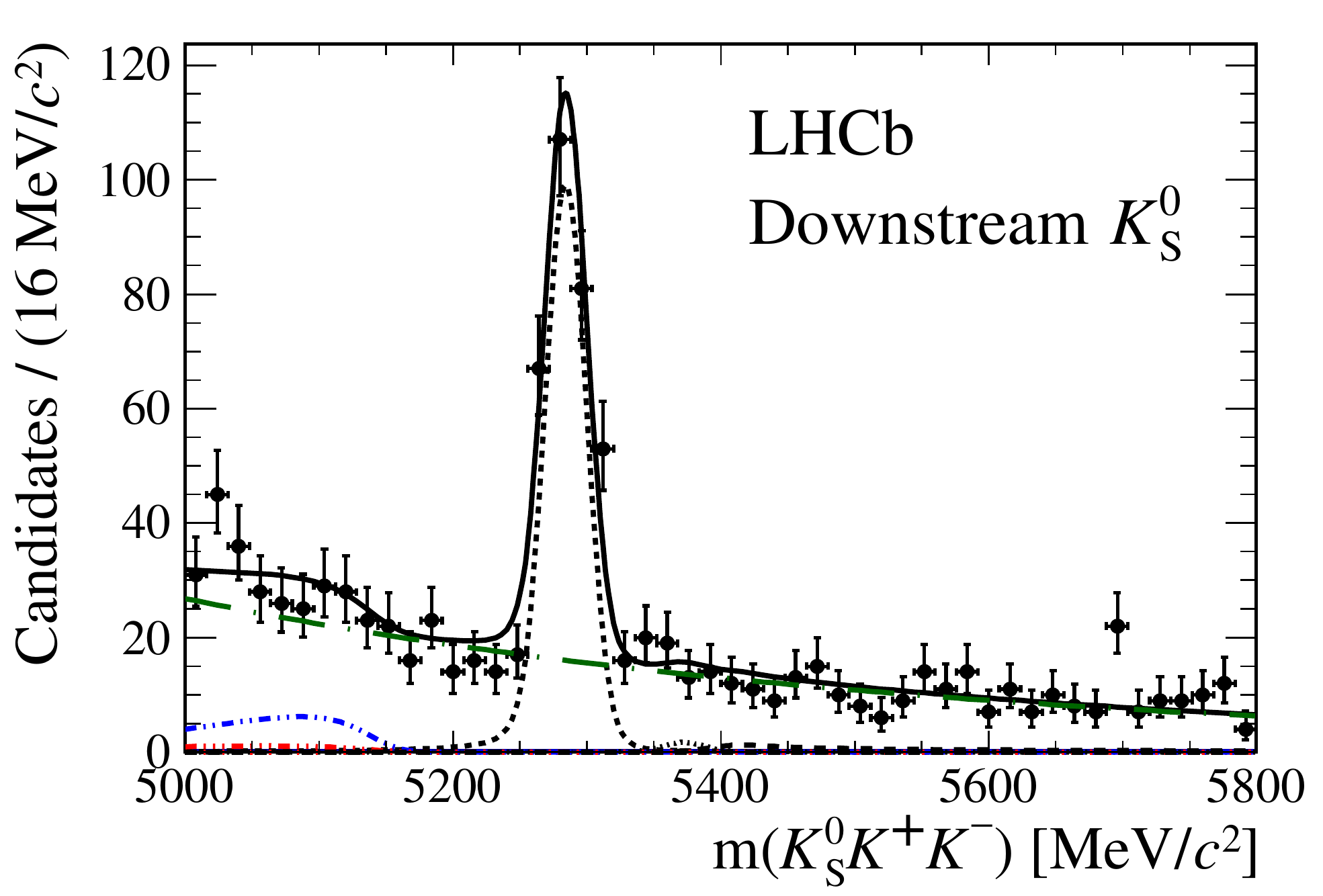}
\includegraphics[width=0.49\textwidth]{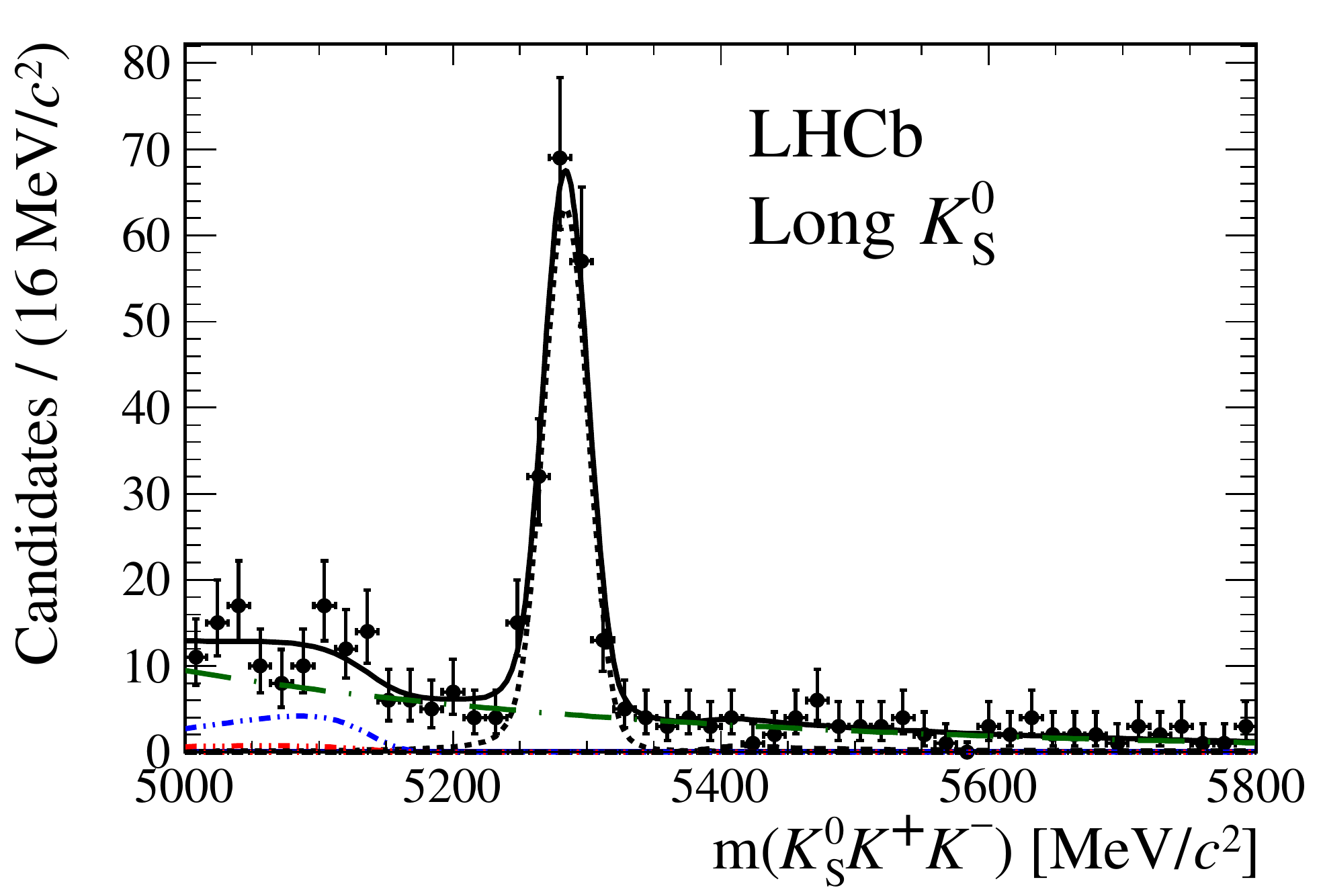}
\includegraphics[width=0.49\textwidth]{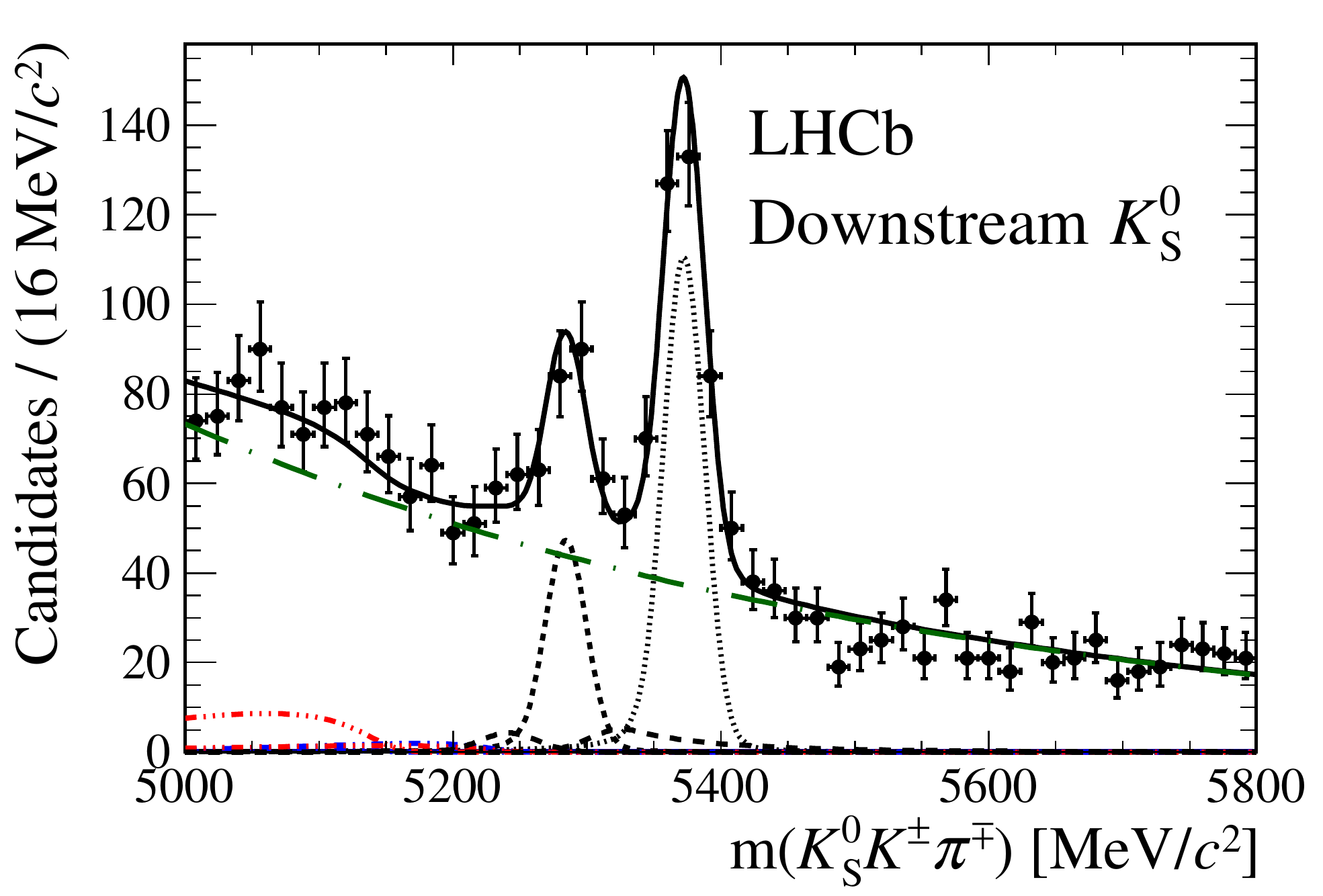}
\includegraphics[width=0.49\textwidth]{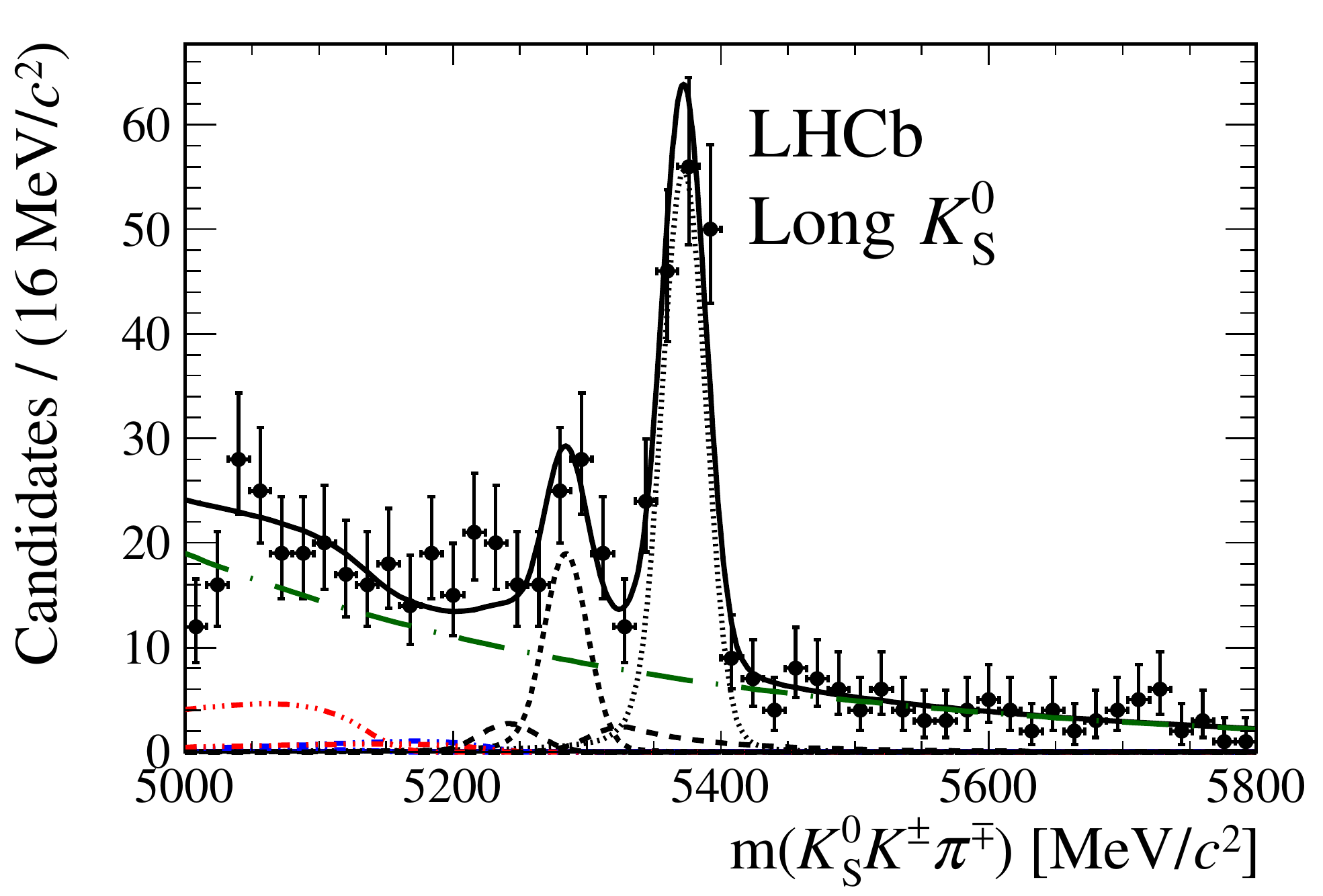}
\includegraphics[width=0.49\textwidth]{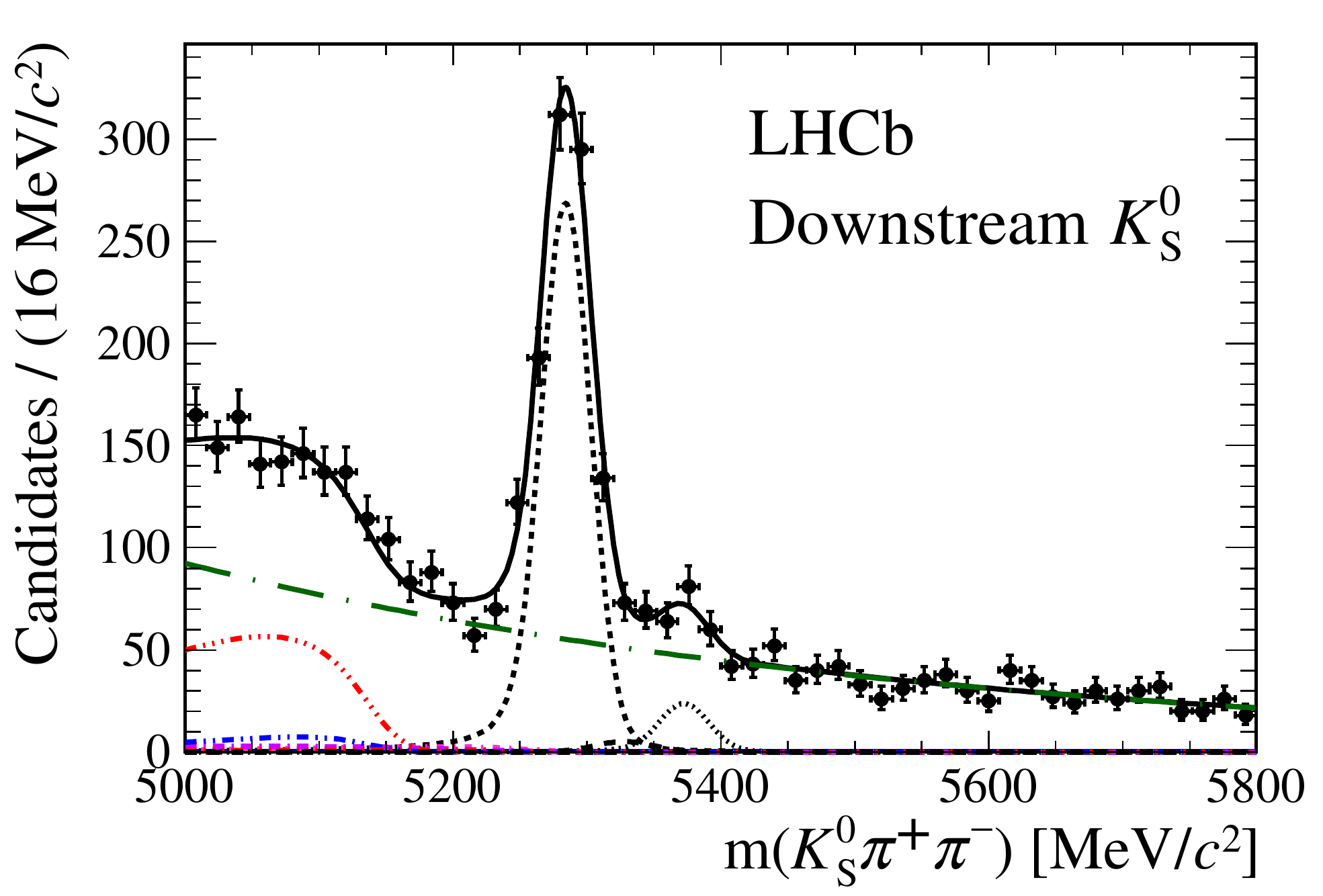}
\includegraphics[width=0.49\textwidth]{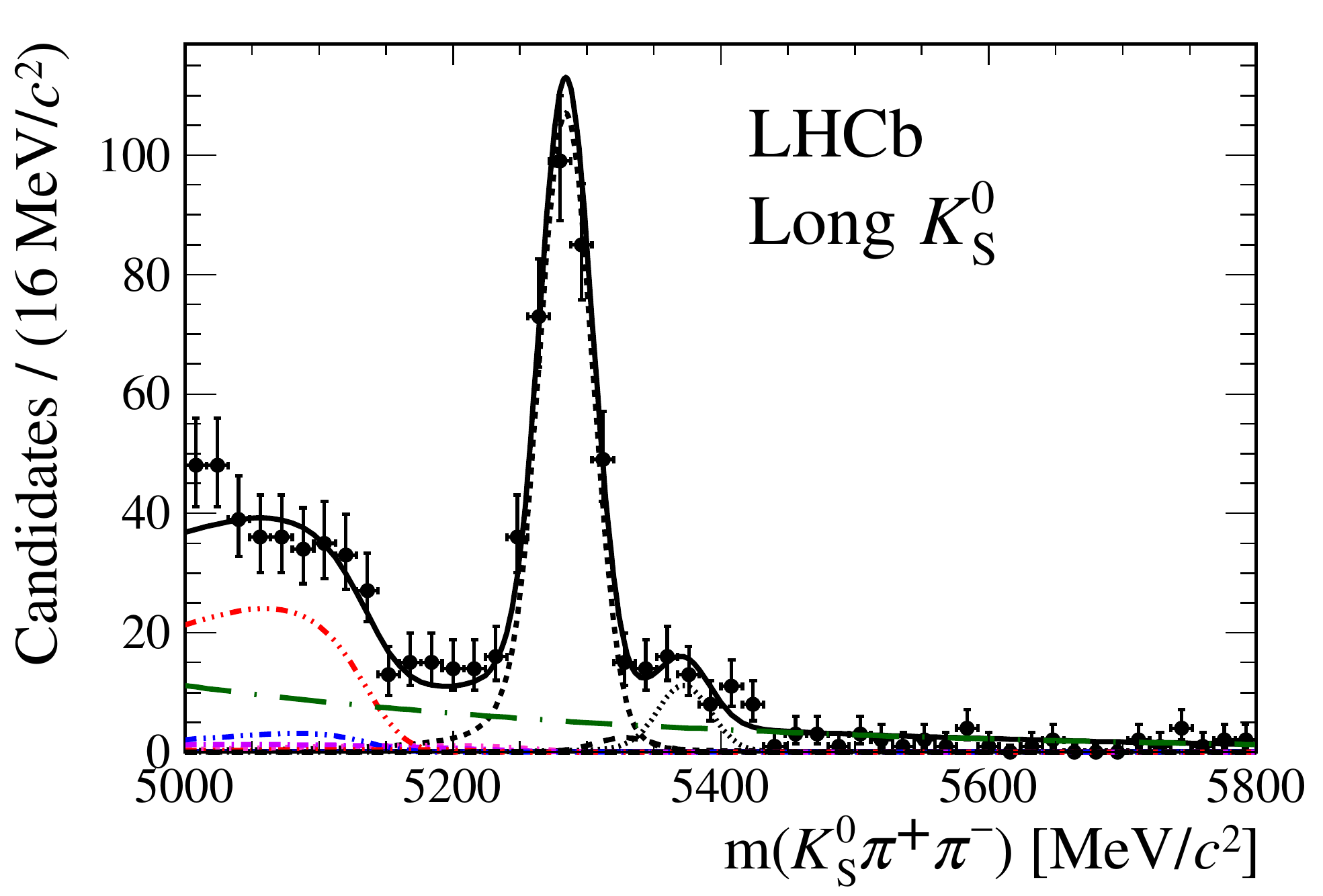}
\caption{
  Invariant mass distributions of (top) \KSKK, (middle) \KSKpi, and (bottom)
  \KSpipi candidate events, with the loose selection for (left) Downstream and 
  (right) Long \KS reconstruction categories.
  In each plot, the total fit model is overlaid (solid black line) on the data
  points.
  The signal components are the black short-dashed or dotted lines,
  while cross-feed decays are the black dashed lines close to the signal peaks.
  The combinatorial background contribution is the green long-dash dotted
  line.
  Partially reconstructed contributions from various sources are also shown.
}
\label{fig:KShh-BR}
\end{figure}

\begin{figure}[htb]
\centering
\includegraphics[width=0.49\textwidth]{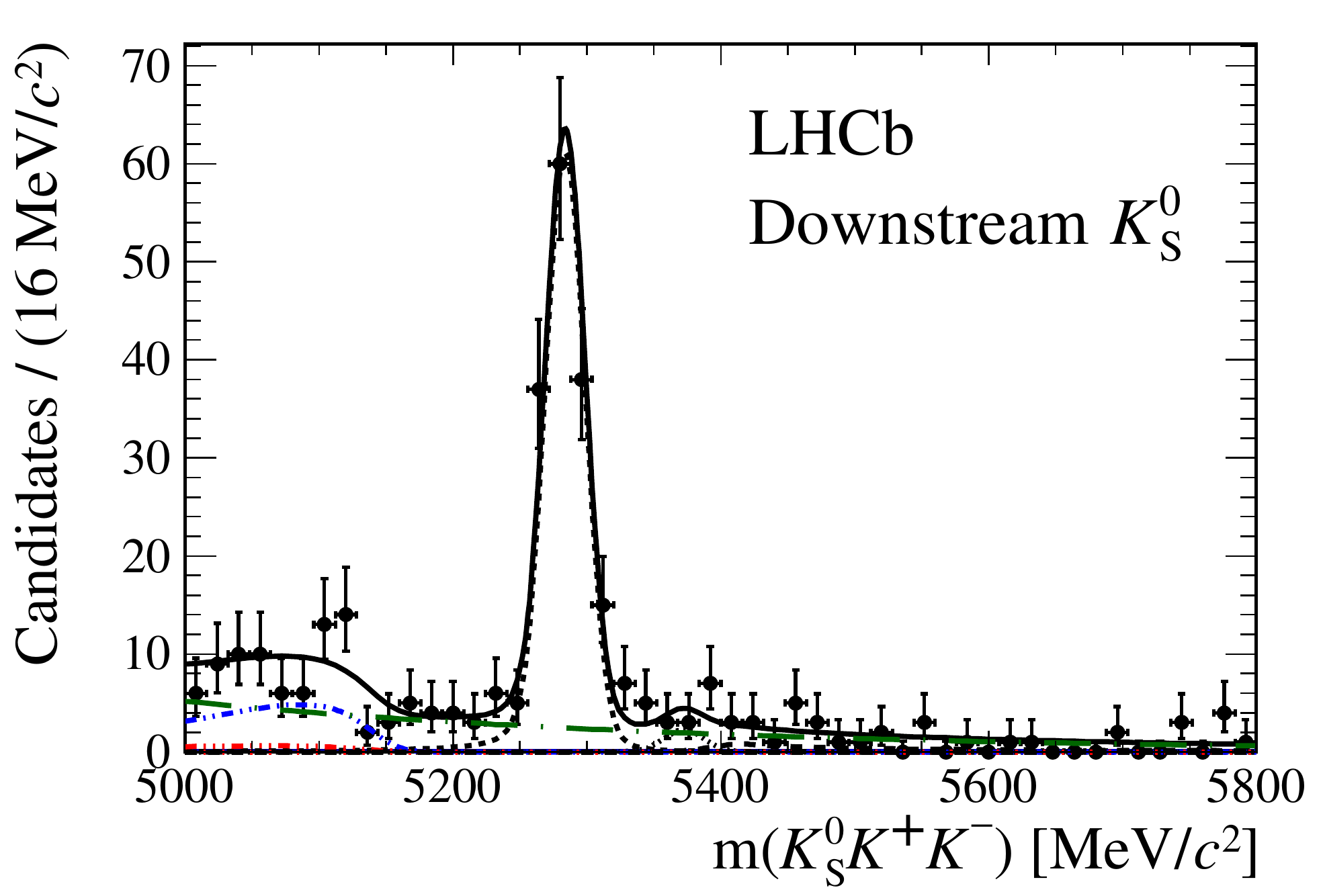}
\includegraphics[width=0.49\textwidth]{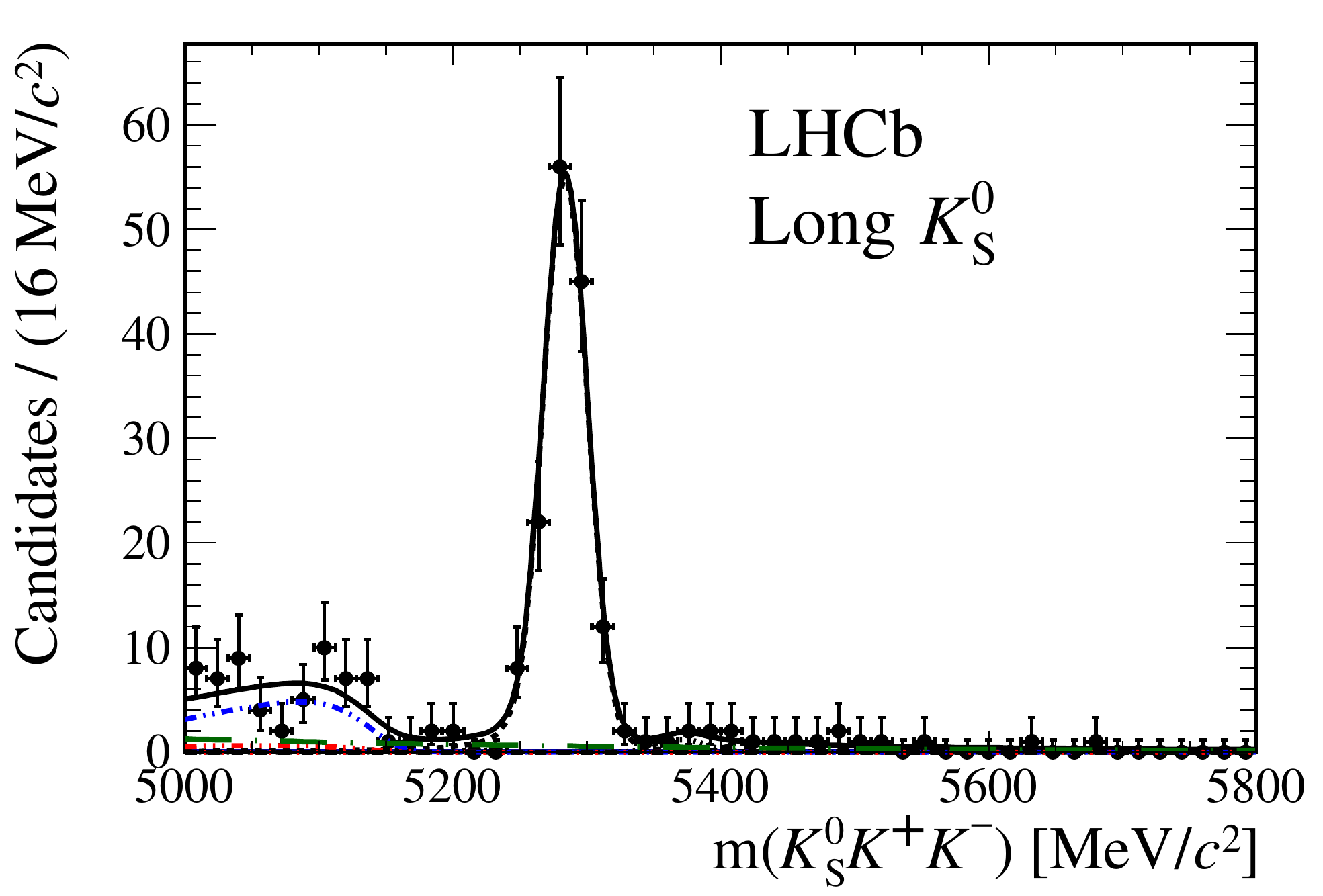}
\includegraphics[width=0.49\textwidth]{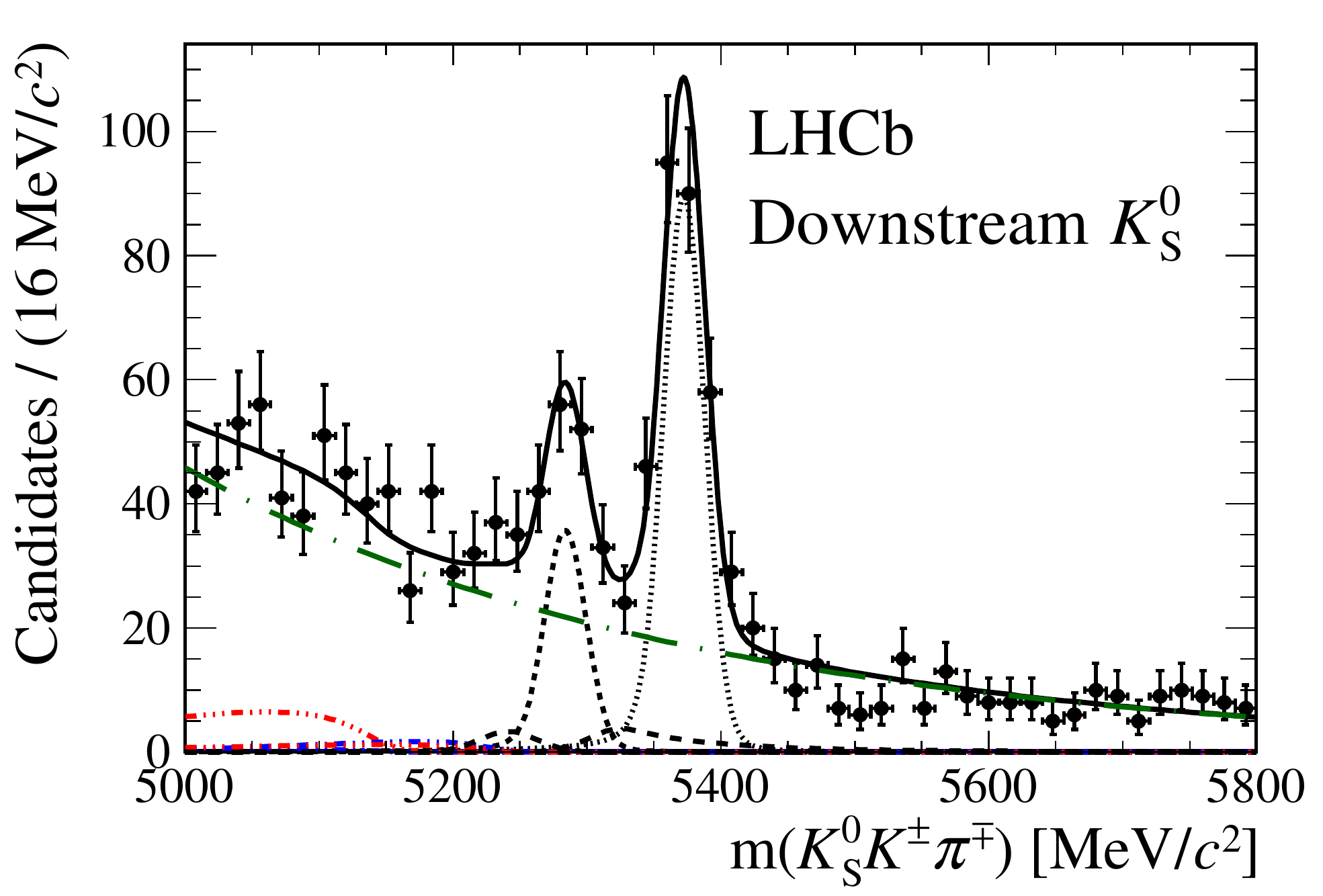}
\includegraphics[width=0.49\textwidth]{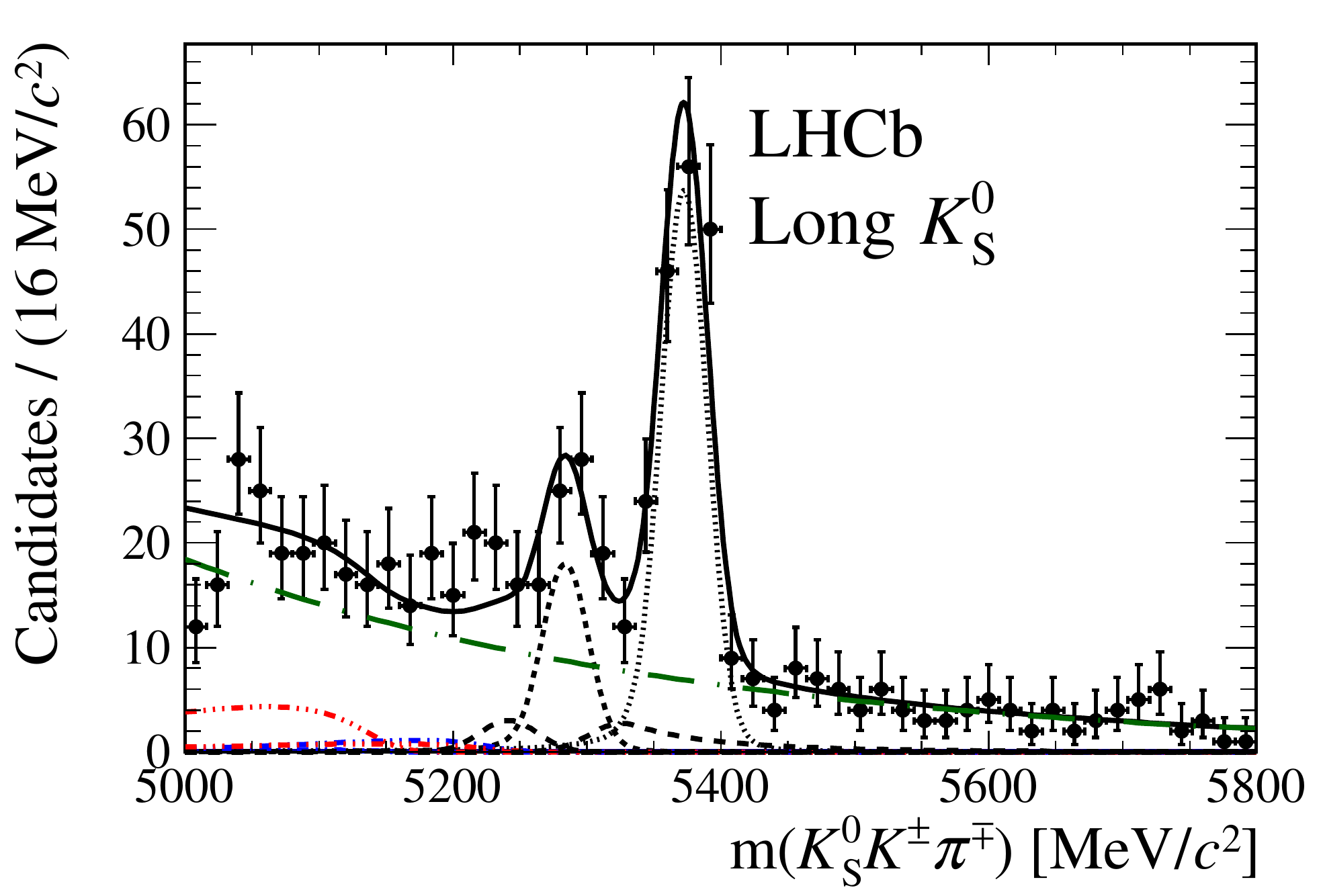}
\includegraphics[width=0.49\textwidth]{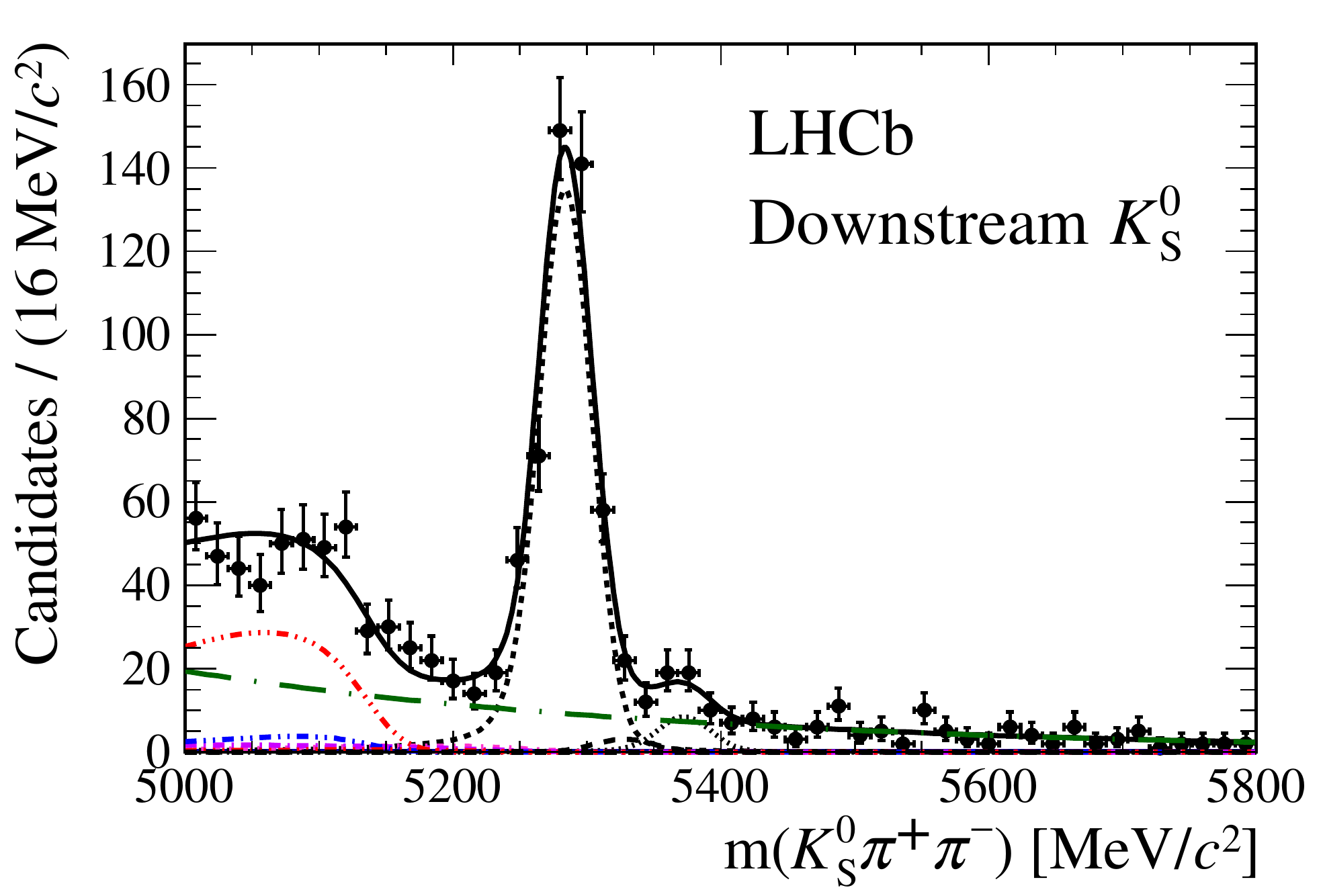}
\includegraphics[width=0.49\textwidth]{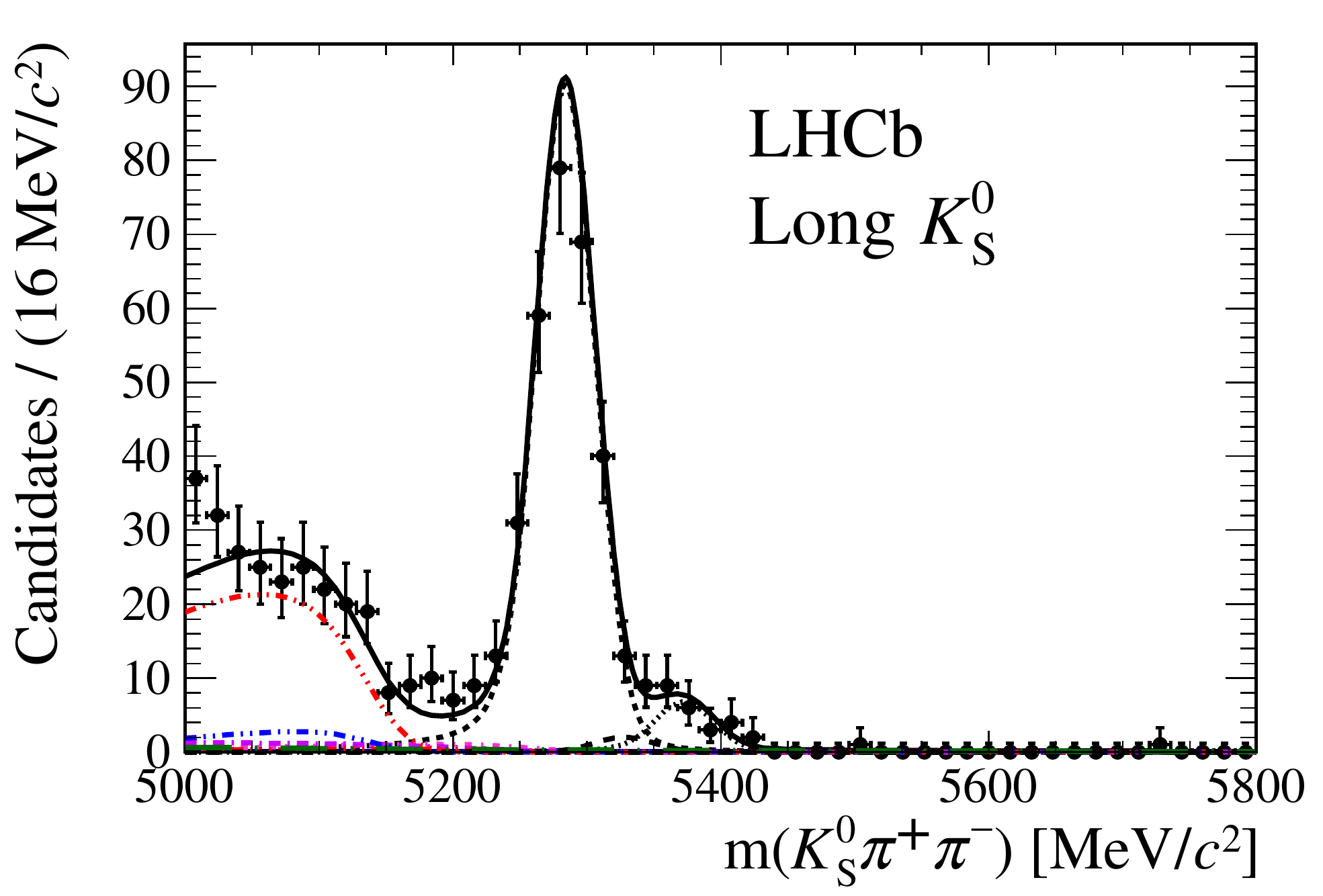}
\caption{
  Invariant mass distributions of (top) \KSKK, (middle) \KSKpi, and (bottom)
  \KSpipi candidate events, with the tight selection for (left) Downstream and 
  (right) Long \KS reconstruction categories.
  In each plot, the total fit model is overlaid (solid black line) on the data
  points.
  The signal components are the black short-dashed or dotted lines,
  while cross-feed decays are the black dashed lines close to the signal peaks.
  The combinatorial background contribution is the green long-dash dotted
  line.
  Partially reconstructed contributions from various sources are also shown.
}
\label{fig:KShh-OBS}
\end{figure}

Figures~\ref{fig:KShh-BR} and~\ref{fig:KShh-OBS} show the results of the fits
to the \BdsToKShh candidate events when the selection is applied for the
favoured modes and suppressed modes, respectively.
The decay \BsToKSKpi is unambiguously observed and the \babar observation of
\BdToKSKpi~\cite{delAmoSanchez:2010ur} is confirmed.
The decay \BsToKSpipi is observed for the first time with a significance of
$5.9\,\sigma$, while no significant evidence is obtained for the decay \BsToKSKK.

The branching fractions of all the modes are measured with respect to
\BdToKSpipi, for which the world average branching fraction is
$(2.48 \pm 0.10) \times 10^{-5}$~\cite{Beringer:2012zz}.
The ratios of branching fractions are determined to be
\begin{eqnarray}
\nonumber
\frac{\Br{\BdToKSKpi}} {\Br{\BdToKSpipi}}  & = & 0.128 \pm 0.017 \; {\rm(stat.)} \; \pm 0.009 \; ({\rm syst.}) \,,\\ 
\nonumber                                              
\frac{\Br{\BdToKSKK}}  {\Br{\BdToKSpipi}}  & = & 0.385 \pm 0.031 \; {\rm(stat.)} \; \pm 0.023 \; ({\rm syst.}) \,,\\
\nonumber                                              
\frac{\Br{\BsToKSpipi}}{\Br{\BdToKSpipi}}  & = & 0.29\phantom{0} \pm 0.06\phantom{0} \; {\rm(stat.)} \; \pm  0.03\phantom{0} \; ({\rm syst.}) \; \pm 0.02\phantom{0} \; (f_s/f_d) \,, \\
\nonumber  
\frac{\Br{\BsToKSKpi}} {\Br{\BdToKSpipi}}  & = & 1.48\phantom{0} \pm 0.12\phantom{0} \; {\rm(stat.)} \; \pm  0.08\phantom{0} \; ({\rm syst.}) \; \pm 0.12\phantom{0} \; (f_s/f_d) \,,\\
\nonumber    
\frac{\Br{\BsToKSKK}}{\Br{\BdToKSpipi}}  &\in& [0.004;0.068]  \; {\rm at \;\; 90\% \; CL}  \,,
\end{eqnarray}
where $f_\squark/f_\dquark$ refers to the uncertainty on the ratio of
hadronisation fractions of the \bquark quark to \Bs and \Bd
mesons~\cite{Aaij:2013qqa}.

\clearpage

\section{Summary}

A review of recent results of the analyses of charmless three-body decays of
\bquark-hadrons has been presented.
With the \B-factories exploiting their final datasets and LHCb starting to
analyse the 2\invfb 2012 data sample there should be many more interesting
results to come in the near future, both in \B meson decays and in the almost
completely unexplored territory of the decays of the \Lb and other
\bquark-baryons.

\Acknowledgments

Work supported by the European Research Council under FP7
and by the United Kingdom's Science and Technology Facilities Council.


\begin{thebibliography}{99}

%%
%%  bibliographic items can be constructed using the LaTeX format in SPIRES:
%%    see    http://www.slac.stanford.edu/spires/hep/latex.html
%%  SPIRES will also supply the CITATION line information; please include it.
%%

\bibitem{Aaij:2013iua} 
  R.~Aaij {\it et al.}  (LHCb Collaboration),
  %``First observation of $CP$ violation in the decays of $B^0_s$ mesons,''
  Phys.\ Rev.\ Lett.\  {\bf 110}, 221601 (2013),
  arXiv:1304.6173 [hep-ex].
  %%CITATION = ARXIV:1304.6173;%%

\bibitem{Gronau:2000zy} 
  M.~Gronau,
  %``U spin symmetry in charmless \B decays,''
  Phys.\ Lett.\ B {\bf 492}, 297 (2000),
  hep-ph/0008292.
  %%CITATION = HEP-PH/0008292;%%

\bibitem{Latham:2008zs} 
  T.~Latham and T.~Gershon,
  %``A Method to Measure $\cos(2\beta)$ Using Time-Dependent Dalitz Plot Analysis of $\Bz \to \D_\CP \pip \pim$,''
  J.\ Phys.\ G {\bf 36}, 025006 (2009),
  arXiv:0809.0872 [hep-ph].
  %%CITATION = ARXIV:0809.0872;%%

\bibitem{Garmash:2005rv} 
  A.~Garmash {\it et al.}  (\belle Collaboration),
  %``Evidence for large direct \CP violation in $\Bpm \to \rho(770)^0 \Kpm$ from analysis of the three-body charmless $\Bpm \to \Kpm \pipm \pimp$ decay,''
  Phys.\ Rev.\ Lett.\  {\bf 96}, 251803 (2006),
  hep-ex/0512066.
  %%CITATION = HEP-EX/0512066;%%

\bibitem{Aubert:2008bj} 
  B.~Aubert {\it et al.}  (\babar Collaboration),
  %``Evidence for Direct \CP Violation from Dalitz-plot analysis of $B^\pm \to K^\pm \pi^\mp \pi^\pm$,''
  Phys.\ Rev.\ D {\bf 78}, 012004 (2008),
  arXiv:0803.4451 [hep-ex].
  %%CITATION = ARXIV:0803.4451;%%

\bibitem{Lees:2012kxa} 
  J.~P.~Lees {\it et al.}  (\babar Collaboration),
  %``Study of \CP violation in Dalitz-plot analyses of $\Bz \to \Kp\Km\KS$, $\Bp \to \Kp\Km\Kp$, and $\Bp \to \KS\KS\Kp$,''
  Phys.\ Rev.\ D {\bf 85}, 112010 (2012),
  arXiv:1201.5897 [hep-ex].
  %%CITATION = ARXIV:1201.5897;%%

\bibitem{Aaij:2013sfa} 
  R.~Aaij {\it et al.}  (LHCb Collaboration),
  %``Measurement of CP violation in the phase space of $B^{\pm} \to K^{\pm} \pi^{+} \pi^{-}$ and $B^{\pm} \to K^{\pm} K^{+} K^{-}$ decays,''
  Phys.\ Rev.\ Lett.\  {\bf 111}, 101801 (2013),
  arXiv:1306.1246 [hep-ex].
  %%CITATION = ARXIV:1306.1246;%%

\bibitem{Alves:2008zz} 
  A.~A.~Alves, Jr. {\it et al.}  (LHCb Collaboration),
  %``The LHCb Detector at the LHC,''
  JINST {\bf 3}, S08005 (2008).
  %%CITATION = JINST,3,S08005;%%

\bibitem{Beringer:2012zz} 
  J.~Beringer {\it et al.}  (Particle Data Group),
  %``Review of Particle Physics (RPP),''
  Phys.\ Rev.\ D {\bf 86}, 010001 (2012).
  %%CITATION = PHRVA,D86,010001;%%

\bibitem{Lees:2013ngt} 
  J.~P.~Lees {\it et al.}  (\babar Collaboration),
  %``Study of the $K^+ K^-$ invariant-mass dependence of \CP asymmetry in $B^+ \rightarrow K^+ K^- K^+$ decays,''
  arXiv:1305.4218 [hep-ex].
  %%CITATION = ARXIV:1305.4218;%%

\bibitem{LHCb-CONF-2012-028}
  R.~Aaij {\it et al.}  (LHCb Collaboration),
  %``Evidence for \CP violation in $B \to KK\pi$ and $B \to \pi\pi\pi$ decays,''
  LHCb-CONF-2012-028.

\bibitem{Aaij:2013fla} 
  R.~Aaij {\it et al.}  (LHCb Collaboration),
  %``Studies of the decays $B^+ \to p \bar p h^+$ and observation of $B^+ \to \kern 0.1em\bar{\kern -0.1em\Lambda}(1520)p$,''
  Phys.\ Rev.\ D {\bf 88}, 052015 (2013),
  arXiv:1307.6165 [hep-ex].
  %%CITATION = ARXIV:1307.6165;%%

\bibitem{Pivk:2004ty} 
  M.~Pivk and F.~R.~Le Diberder,
  %``SPlot: A Statistical tool to unfold data distributions,''
  Nucl.\ Instrum.\ Meth.\ A {\bf 555}, 356 (2005),
  physics/0402083 [physics.data-an].
  %%CITATION = PHYSICS/0402083;%%

\bibitem{Aaij:2013rha} 
  R.~Aaij {\it et al.}  (LHCb Collaboration),
  %``Measurements of the branching fractions of $B^{+} \to p \bar p K^{+}$ decays,''
  Eur.\ Phys.\ J.\ C {\bf 73}, 2462 (2013),
  arXiv:1303.7133 [hep-ex].
  %%CITATION = ARXIV:1303.7133;%%

\bibitem{Gaur:2013uou} 
  V.~Gaur {\it et al.}  (\belle Collaboration),
  %``Evidence for the decay $B^0 \to K^+ K^- \pi^0,''
  Phys.\ Rev.\ D {\bf 87}, 091101 (2013),
  arXiv:1304.5312 [hep-ex].
  %%CITATION = ARXIV:1304.5312;%%

\bibitem{Snyder:1993mx} 
  A.~E.~Snyder and H.~R.~Quinn,
  %``Measuring \CP asymmetry in $\B \to \rho \pi$ decays without ambiguities,''
  Phys.\ Rev.\ D {\bf 48}, 2139 (1993).
  %%CITATION = PHRVA,D48,2139;%%

\bibitem{Lees:2013nwa} 
  J.~P.~Lees {\it et al.}  (\babar Collaboration),
  %``Measurement of \CP-violating asymmetries in $B^0 \to (\rho \pi)^0$ decays using a time-dependent Dalitz plot analysis,''
  Phys.\ Rev.\ D {\bf 88}, 012003 (2013),
  arXiv:1304.3503 [hep-ex].
  %%CITATION = ARXIV:1304.3503;%%

\bibitem{Aaij:2013uta} 
  R.~Aaij {\it et al.}  [LHCb Collaboration],
  %``Study of $B_{\scriptscriptstyle (s)}^0 \to K_{\rm \scriptscriptstyle S}^0 h^{+} h^{\prime -}$ decays with first observation of $B_{\scriptscriptstyle s}^0 \to K_{\rm \scriptscriptstyle S}^0 K^{\pm} \pi^{\mp}$ and $B_{\scriptscriptstyle s}^0 \to K_{\rm \scriptscriptstyle S}^0 \pi^{+} \pi^{-}$,''
  to appear in JHEP,
  arXiv:1307.7648 [hep-ex].
  %%CITATION = ARXIV:1307.7648;%%

\bibitem{delAmoSanchez:2010ur} 
  P.~del Amo Sanchez {\it et al.}  (\babar Collaboration),
  %``Observation of the Rare Decay $\Bz \to \KS\Kpm\pimp$,''
  Phys.\ Rev.\ D {\bf 82}, 031101 (2010),
  arXiv:1003.0640 [hep-ex].
  %%CITATION = ARXIV:1003.0640;%%

\bibitem{Aaij:2013qqa} 
  R.~Aaij {\it et al.}  (LHCb Collaboration),
  %``Measurement of the fragmentation fraction ratio $f_{s}/f_{d}$ and its dependence on $B$ meson kinematics,''
  JHEP {\bf 1304}, 001 (2013),
  arXiv:1301.5286 [hep-ex].
  %%CITATION = ARXIV:1301.5286;%%

\end{thebibliography}
\end{document}